\PassOptionsToPackage{dvipsnames}{xcolor}
\documentclass[conference]{IEEEtran}
\IEEEoverridecommandlockouts
\usepackage{cite}
\usepackage{amssymb,amsmath,amsthm, amsfonts}
\usepackage{algorithmic}
\usepackage{mathtools}
\usepackage{float}
\usepackage{graphicx}
\graphicspath{{pics/}}
\usepackage{textcomp}
\usepackage{xcolor}
\usepackage{url}
\usepackage[shortlabels]{enumitem}
\newlist{nenum}{enumerate}{10}
\setlist[nenum]{label*=\arabic*.}
\def\BibTeX{{\rm B\kern-.05em{\sc i\kern-.025em b}\kern-.08em
    T\kern-.1667em\lower.7ex\hbox{E}\kern-.125emX}}

\usepackage{comment}
\usepackage{subcaption}
\usepackage{listings}
\usepackage{float}
\usepackage{mathtools, nccmath}

\begin{document}

\newtheorem{theorem}{Theorem}
\newtheorem{definition}{Definition}
\newtheorem{lemma}{Lemma}

\title{{Wait-free} Trees with Asymptotically-Efficient Range Queries}

\author{\IEEEauthorblockN{Ilya Kokorin}
\IEEEauthorblockA{ITMO University, VK}
\and
\IEEEauthorblockN{Dan Alistarh}
\IEEEauthorblockA{IST Austria}
\and
\IEEEauthorblockN{Vitaly Aksenov}
\IEEEauthorblockA{City, University of London}
% \and
% \and
% \and
}

\maketitle

\begin{abstract}
Tree data structures, such as red-black trees, quad trees, treaps, or tries, are fundamental tools in computer science. A classical problem in concurrency is to obtain expressive, efficient, and scalable versions of practical tree data structures. 
We are interested in concurrent trees supporting \emph{range queries}, i.e., queries that involve multiple consecutive data items. 
Existing implementations with this capability can list keys in a specific range, but do not support \emph{aggregate range queries}: for instance, if we want to calculate the number of keys in a range, the only choice is to retrieve a whole list and return its size. This is suboptimal: in the sequential setting, one can augment a balanced search tree with counters and, consequently, perform these aggregate requests in logarithmic rather than linear time.

In this paper, we propose a generic approach to implement a broad class of range queries on concurrent trees in a way that is wait-free, asymptotically efficient, and practically scalable. 
The key idea is a new mechanism for maintaining metadata concurrently at tree nodes, which can be seen as a wait-free variant of hand-over-hand locking (which we call \emph{hand-over-hand helping}).
We implement, test, and benchmark a balanced binary search tree with {wait-free} \texttt{insert}, \texttt{delete}, \texttt{contains}, and \texttt{count} operations, returning the number of keys in a given range which validates the expected speedups because of our method in practice. 
\end{abstract}


\section{Introduction}
\label{introduction-section}

Tree data structures are ubiquitous in computer science, due to their high expressive power and practical versatility. 
For instance, in databases, index trees allow searching for an indexed key faster than traversing through all the elements. Typically, such index is implemented as B-tree~\cite{bayer2012r,comer1979ubiquitous,graefe2011modern}, although alternate implementations are possible, such as the red-black tree~\cite{guibas1978dichromatic}, or the splay tree~\cite{sleator1985self}.
Moreover, one could use quad trees~\cite{de2000computational} to store and retrieve a collection of points in a plane, or tries~\cite{bodon2003trie} for fast prefix matching in strings.

%We start with the simple definition of a generic tree. \emph{Tree} is a data structure that consists of a set of \emph{nodes}. Each node can be connected to many \emph{child} nodes, but can have no more than one \emph{parent} node. All nodes, except for the one, called the \emph{root} node, have exactly one parent. Root node does not have a parent. For more information see~\cite{tree-definition}.

In this paper, we are interested in \emph{concurrent} implementations of fundamental tree data structures that combine theoretical and practical efficiency, with  \emph{expressivity} in terms of the class of queries they support efficiently. 
Specifically, we are interested in trees supporting the following types of operations. We call a query, retrieving or modifying a single data item, a \emph{scalar query}; and a query, involving multiple consecutive (by value) data items, a \emph{range query}. 
For example, a search tree can provide the following scalar queries:
\begin{itemize}
    \item \texttt{insert(key)}~--- if \texttt{key} does not exist in the tree, inserts it to the tree, otherwise, leaves the tree unmodified;
    
    \item \texttt{remove(key)}~--- if \texttt{key} exists in the tree, removes it from the tree, otherwise, leaves the tree unmodified;
    
    \item \texttt{contains(key)}~--- returns \texttt{true} if the tree contains \texttt{key}, \texttt{false}, otherwise.
\end{itemize} 

Also, a search tree can provide the following range queries:
\begin{itemize}    
    \item \texttt{collect(min, max)}~--- returns all the keys from the \texttt{[min; max]} interval from the set;

    \item \texttt{count(min, max)}~--- returns the number of keys from the \texttt{[min; max]} interval from the set.
\end{itemize}

In addition, we would like to support aggregate range queries: for example, in a search tree storing key-value pairs, the range query \texttt{range\_add(min, max, delta)} adds \texttt{delta} to all the values corresponding to the keys in a given range \texttt{[min, max]}, whereas the range query \texttt{range\_sum(min, max)} calculates the sum of all values corresponding to the keys in a given range \texttt{[min, max]}.

%Note that for the simplicity of the presentation, we choose a binary tree as our target index.

In this work, in addition to extensively investigated \texttt{collect} query (see e.g., ~\cite{brown2012range,arbel2018harnessing}) we require the index to perform aggregate range queries (e.g., \texttt{count}) in an asymptotically optimal way. For example, we can use such aggregate range queries to find the number of requests to the system in the specified time range from the specified users.

Currently, all existing concurrent trees answer the aggregate range queries in time proportional to the number of elements in the range, i.e., for a count query it works as \texttt{count(min, max) = collect(min, max).length()}.
This is clearly suboptimal: in the sequential setting, augmented search trees can perform such queries in $O(\texttt{height})$ (where \texttt{height} is the height of the tree) which can be exponentially faster for balanced trees.

Now, we overview how to sequentially perform \texttt{count} query in $O(\texttt{height})$ time for a binary search tree.
Note that other aggregate range queries can be implemented similarly.
For each node, we store the number of keys in its subtree.
Then, we start traversing the tree from the root downwards.
When we are in the node \texttt{v} we check three cases.
If the range of keys in the subtree of \texttt{v} lies inside the required range~--- we add the stored size of the subtree of \texttt{v} to the answer.
If it intersects with the required range~--- we go recursively to children, returning the sum of results for \texttt{v.left} and \texttt{v.right}. 
And, finally, if it does not intersect with the required range~--- we stop the call and return zero. (We unroll this recursion in our sequential and concurrent implementations we unroll the recursion.) 
% We will note that, in our sequential and concurrent implementations we unroll the recursion. We provide a more detailed description of the sequential algorithm, as well as the proof of the $O(height)$ time complexity, in Appendix~\ref{sequential-algorithm}.

In this paper, we present a scalable approach that can make any tree data structure support wait-free operations, including asymptotically efficient aggregate range queries with logarithmic amortized time. The main idea 
is that the execution of an operation \texttt{Op} by a process \texttt{P} at node $\texttt{v}$ begins by inserting the descriptor of \texttt{Op} into the root queue \texttt{root.Queue}, and obtaining a 
timestamp. Then, the process \texttt{P} helps to perform all pending operations in the queue, applies itself, and proceeds recursively at the children nodes, applying the same pattern. 
Thus, the process traverses the tree downwards, from the root to the appropriate lower nodes, at which the operation (e.g., an insertion of a new data item, or a removal of an existing one) should be performed.
This method can be seen as a \emph{wait-free} version of the classic \emph{hand-over-hand locking} technique~\cite{herlihy2020art}, where instead of blocking we ask processors to perform work that is preceding them in the queue. 
We name this method as \emph{hand-over-hand helping}.
In the following, we describe this construction in detail, using a binary search tree as a running example.
Finally, we provide a practical implementation of such a tree, supporting \texttt{insert}, \texttt{delete}, \texttt{contains}, and \texttt{count} operations. We validate the fact that our design permits the efficient implementation of various types of range queries while achieving non-trivial scalability.

\subsection{Related work}
\textbf{Lock-based solutions.}
The easiest and the most obvious way to implement a concurrent data structure is to protect a sequential data structure with a \emph{lock} to guarantee mutual exclusion~\cite{lamport2019new}. Such construction is not lock-free (it is not even obstruction-free) and suffers from starvation. Moreover, since a lock allows only one process to work with the data structure at a time, the resulting construction does not scale and its throughput remains low.

\textbf{Linear-time solutions.}
Several papers \cite{arbel2018harnessing, brown2012range, fatourou2019persistent, wei2021constant, basin2017kiwi} address the issue of executing lock-free (and even wait-free, but with lock-free scalar queries) range queries on concurrent trees. However, the aforementioned papers address only the \texttt{collect(min, max)} query, returning the list of keys, located within a range \texttt{[min; max]}. All other range queries are proposed to be implemented on top of the \texttt{collect} query. For example, as we said before, the \texttt{count} query can be implemented as \texttt{count(min, max) = collect(min, max).length()}. 

This approach suffers from a major drawback: the \texttt{collect} query is executed in time proportional to the number of keys in the range. Thus, for wide ranges, such query takes $O(N)$ time where $N$ is the size of the tree: the number of keys in the range is almost equal to the size of the tree. This implementation is not asymptotically efficient: e.g., the \texttt{count} query can be executed in $O(\log N)$ time in a sequential environment using balanced search trees.

Therefore, despite being lock-free, these methods do not guarantee time efficiency, and thus cannot be used.

\textbf{Persistent data structures.}
There exists a solution for efficient aggregate range queries based on persistent data structures~\cite{persistent-data-structures}. Each read-only operation (e.g., \texttt{contains} or \texttt{count}) takes the current version of the data structure and operates on it. Each update operation (e.g., \texttt{insert} or \texttt{remove}) creates a new version of the data structure without modifying the existing one and then tries to replace the old version with the new one using a Compare-And-Swap~\cite{compare-and-swap} (or CAS, for simplicity).
If the CAS succeeds the operation finishes, otherwise the operation restarts from the very beginning.
This approach is called Lock-free Universal Construction~\cite{herlihy2020art} and can 
be applied to any sequential persistent tree.
As an interesting observation, this approach scales even on write-only workloads~\cite{aksenov2023unexpected}.
However, there are at least two drawbacks: 1)~we cannot provide strong fairness guarantees~--- one operation can restart infinitely often if we are not lucky enough; 2)~for an update range query, the majority of computation time will be spent needlessly~--- since unsuccessful CAS makes us retry the whole operation from the very beginning.
For more information about this approach, we point to~\cite{aksenov2023unexpected}.

\textbf{Parallel augmented persistent trees.}
Sun, Ferizovic, and Belloch~\cite{sun2018pam} presented a persistent augmented tree that can serve a batch of operations in parallel using \emph{fork-join} parallelism.
The paper does not propose a method of executing concurrent operations on augmented data structures.
However, we can use various combining techniques~\cite{sun2019supporting,hendler2010flat,aksenov2019parallel} to form large batches of operations from individual concurrent updates.
The main problem with this approach is that the combining techniques increase individual operation latency and, thus, are not acceptable in settings, where low operation latency is required.

\section{Overview of the approach}
\label{algorithm-general-chapter}

%%Before going into details, we want to remind that we explain our approach on a binary search tree.

\subsection{Timestamps invariant}
\label{main-invariant-chapter}

The main problem with the sequential algorithm for an aggregate range query presented in the introduction is that it will be incorrect if running as is in a concurrent environment. Indeed, each update operation (e.g. \texttt{insert} or \texttt{remove}) should modify not only the tree structure, but the augmentation values on the path (e.g., subtree sizes). By that, the augmentation values may become inconsistent with the tree structure.

Therefore, the main purpose of our concurrent solution is to get rid of such situations by ensuring that all operations are executed in a particular order. We enforce a particular execution order by maintaining an operation queue in each node.

\begin{figure}
  \centering
  \caption{Node \texttt{v} has an operations queue with descriptors of three operations: $\texttt{Op}_{\texttt{1}}$, $\texttt{Op}_{\texttt{2}}$ and $\texttt{Op}_{\texttt{3}}$. These three operations should be applied to \texttt{vs} in the order of descriptors in the queue: first $\texttt{Op}_{\texttt{1}}$, then $\texttt{Op}_{\texttt{2}}$, and, finally, $\texttt{Op}_{\texttt{3}}$}
  \label{main-invariant-pic}
  \includegraphics[width=\linewidth]{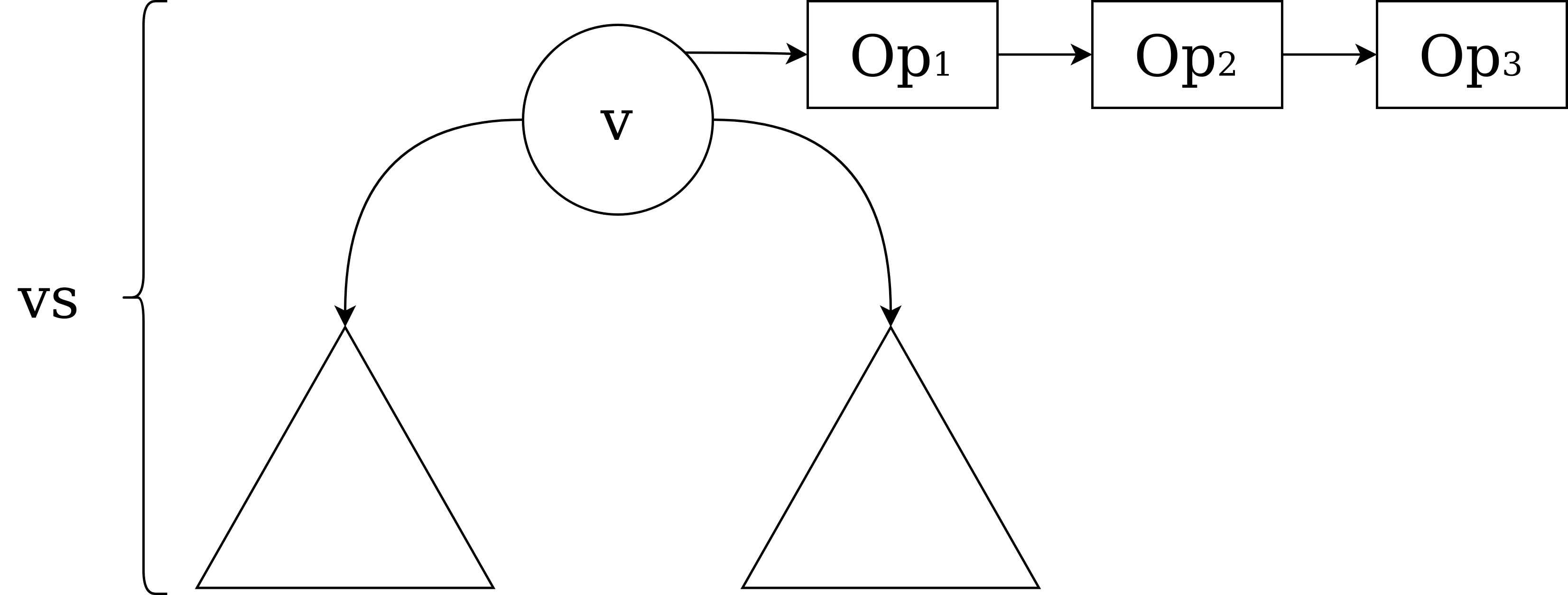}
\end{figure}

Consider an arbitrary node \texttt{v} and its subtree \texttt{vs}. At \texttt{v} we maintain an \emph{operations queue}, that contains descriptors of operations to be applied to \texttt{vs} (Fig.~\ref{main-invariant-pic}). These operations can, for example, insert a key to \texttt{vs} or remove a key from \texttt{vs}. We maintain the following invariant: operations should be applied to \texttt{vs} in the order, their descriptors were added to \texttt{v} queue.

Note that the aforementioned invariant can be applied to the root node too: indeed, since the whole tree is just the subtree of the root operations should be applied to the tree in the order their descriptors were added to the operations queue in the root. Thus, the order, in which operation descriptors are added to the queue in the root, is exactly the \emph{linearization order}.

Thus, we may use the operations queue at the root to allocate timestamps for operations. A timestamp allocation mechanism should provide the following guarantee: if a descriptor of operation \texttt{A} was added to the root queue before a descriptor of operation \texttt{B}, then \texttt{timestamp(A) $<$ timestamp(B)} should hold. We explain how to achieve it in Section~\ref{queue-section}.
%Please note that the timestamps give the correct linearization order.
We store the timestamp of an operation in the corresponding descriptor, i.e., \texttt{descriptor.Timestamp} field.

\begin{figure}
  \centering
  \caption{Two operations can be executed in arbitrary order (even in parallel) if they operate on different subtrees, since on lover tree levels they are placed to different queues}
  \label{parallelism-how-pic}
  \includegraphics[width=\linewidth]{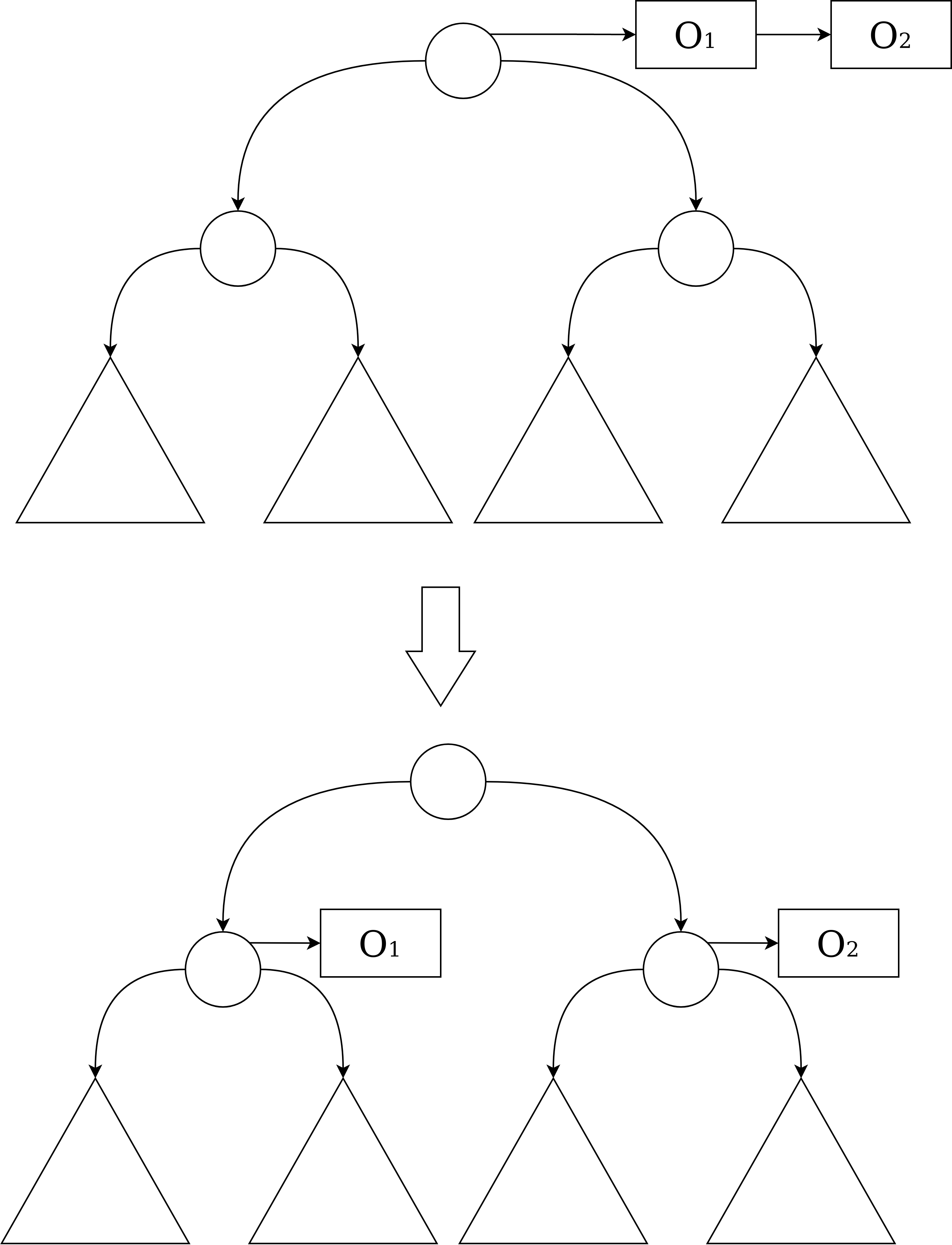}
\end{figure}

As was stated before, operations should be applied to the tree in the order, their descriptors were added to the root descriptor queue. Therefore, one can wonder: how can we achieve parallelism, while linearizing all operations via the root queue? Note that there is no parallelism only in the queue in the root. Lower by the tree, two operations (even the modifying ones, e.g., two \texttt{insert}s) may be executed in parallel if they are executed on different subtrees, since on lower tree levels their descriptors will be placed to different operation queues (Fig.~\ref{parallelism-how-pic}). 

% Note that a particular execution order, determined by the queue, is enforced only in the root node~--- in the root node we execute $O_1$ before starting executing $O_2$. At lower tree levels, we do not enforce a particular execution order. This can be achieved because at lower tree levels $O_1$ and $O_2$ do not conflict with each other anymore, since they operate on different subtrees. Thus, our logical order (in which $O_1$ precedes $O_2$) does not require us to enforce a particular physical execution order on these operations, i.e., executing $O_1$ before starting executing $O_2$.

\subsection{Operation execution: overview}
\label{operation-execution-chapter}

For simplicity of the overview, we consider only unbalanced trees for now. If we want to make our tree balanced, we can adapt the subtree rebuilding approach (we provide a detailed description in Section~\ref{balancing-section}).
The study of other concurrent balancing strategies we leave for the future work.

The execution of an operation \texttt{Op} by a process \texttt{P} (we call such process \texttt{P} the \emph{initiator} process) begins with inserting the descriptor of \texttt{Op} into the root queue and obtaining \texttt{Op} timestamp. In Section~\ref{queue-section}, we describe, how the root queue with timestamp allocation may be implemented.

After that, the initiator process starts traversing the tree downwards, from the root to the appropriate lower nodes, at which the operation (e.g., an insertion of a new data item, or a removal of an existing one) should be performed.

In each visited node \texttt{v} some additional actions should be performed in order to execute \texttt{Op} properly. For example, during the \texttt{count} query the size of \texttt{v}'s subtree can be added to the result, and during \texttt{insert} or \texttt{remove} operations pointers to \texttt{v}'s children and \texttt{v}'s subtree size can be changed. We call the process of performing these necessary actions~--- an \emph{execution of operation \texttt{Op} in node \texttt{v}}.

\begin{figure}[H]
  \centering
  \caption{Process \texttt{P} has to wait before executing \texttt{Op} in node \texttt{v}, since only the operation $\texttt{D}_{\texttt{0}}$, corresponding to the descriptor at the head of \texttt{v} queue, can be executed right now in \texttt{v}.}
  \label{operation-execution-waiting-pic}
  \includegraphics[width=\linewidth]{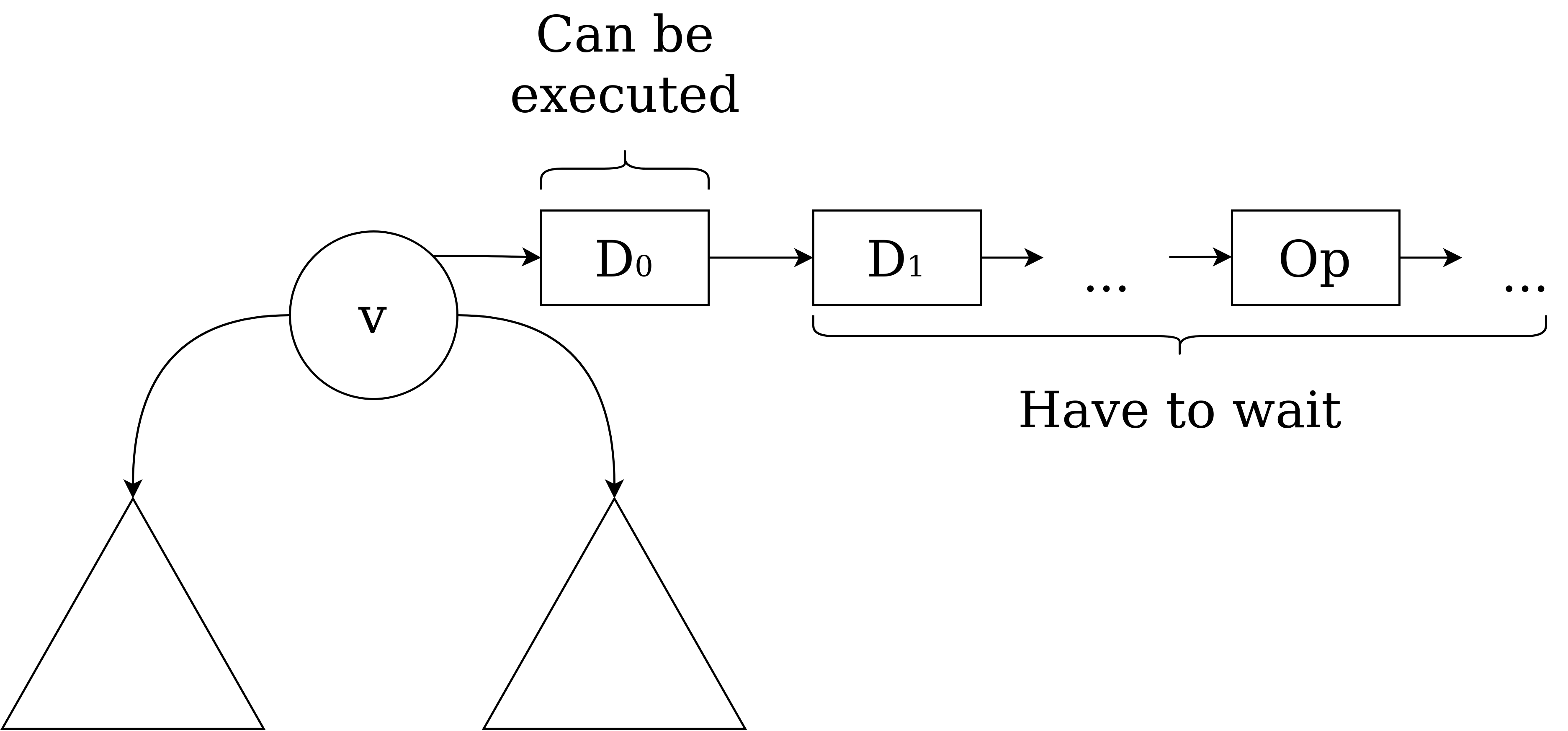}
\end{figure}

As stated in the previous subsection, operations should be applied to \texttt{v}'s subtree in the order their descriptors appear in \texttt{v}'s operations queue.
%(as stated in Section~\ref{main-invariant-chapter}, it equals the order in which operation descriptors appear in the \texttt{v.Queue}).
%operation descriptors are inserted to \texttt{v}'s queue. 
Thus, if the descriptor of \texttt{Op} is not located at the head of \texttt{v}'s queue the initiator process \texttt{P} has to wait before executing \texttt{Op} in node \texttt{v} (Fig.~\ref{operation-execution-waiting-pic}). The execution of \texttt{Op} in node \texttt{v} cannot begin until execution of all the preceding operations in node \texttt{v} is finished.

To make the algorithm wait-free we use the \emph{helping} mechanism (e.g.,~\cite{michael1996simple,harris2002practical}): instead of merely waiting for the \texttt{Op} descriptor to move to the head of \texttt{v} queue, \texttt{P} helps executing in node \texttt{v} the operation from the head of the queue~--- $\texttt{D}_{\texttt{0}}$ in the example above. Thus, even if the initiator process of $\texttt{D}_{\texttt{0}}$ is suspended, the system still makes progress. 

As discussed later, while helping to execute operations $\texttt{D}_{\texttt{0}}, \texttt{D}_{\texttt{1}}, \ldots$ in node \texttt{v} the process \texttt{P} removes descriptors of these operations from the head of \texttt{v}'s queue and inserts them to queues of appropriate \texttt{v}'s children. Thus, while helping other processes execute their initiated operations in \texttt{v}, \texttt{P} moves \texttt{Op} descriptor closer to the head of \texttt{v} queue.
Once \texttt{P} helped all preceding operations to finish their execution in node \texttt{v}, it can finally execute its operation \texttt{Op} in \texttt{v} (note that some other process may help executing \texttt{Op} in \texttt{v}, just like \texttt{P} previously helped executing $\texttt{D}_\texttt{0}$ in \texttt{v}).

The process of executing an operation \texttt{Op} in a node \texttt{v} consists of the following actions:
\begin{nenum}
    \item Determine the set of child nodes \texttt{C}, in which \texttt{Op} execution should continue. 
    
    For example, an execution of the \texttt{count} query on a binary search tree may continue in either single child or both children, as explained in Section~\ref{introduction-section}.
    %: consider the explanations in Section~\ref{sequential-algorithm}~--- the execution continues in both children iff node \texttt{v} is a \emph{fork node} and in a single child (either left or right) otherwise. In contrast, the \texttt{insert} operation should always be continued in a single child, since any key should be stored in exactly one leaf of the tree.
    
    \item For each child \texttt{c} from the set \texttt{C}:
    \begin{nenum}
        \item Modify the state of \texttt{c} (e.g., a size of \texttt{c}'s subtree), if necessary;
        
        \item Try to insert \texttt{Op} descriptor to the end of \texttt{c}'s operations queue, thus allowing \texttt{Op} to continue its execution at lower levels of the tree.
    \end{nenum}
    
    \item Remove \texttt{Op} descriptor from the head of \texttt{v}'s queue.
\end{nenum}

Note, that during the execution of operation \texttt{Op} in node \texttt{v} the said operation only modifies states of \texttt{v}'s children, not \texttt{v} itself. Thus, no operation can ever modify the root state, since the root is not a child of some other node. We overcome that issue by the introduction of the \emph{fictive root}. This fictive root does not contain any state and has only one child~--- the real tree root. The only purpose of the fictive root is to allow operations to modify the state of the real root.
The state of the real root can be modified by operation \texttt{Op} while \texttt{Op} is being executed in the fictive root, since the real root is the child of the fictive root.

%\begin{figure}[H]
%   \centering
%   \caption{The fictive root of the tree with no state and the only child: the real root}
%   \label{fictive-root-pic}
%   \includegraphics[width=0.5\linewidth]{pics/fictive-root.png}
%\end{figure}

In Section~\ref{no-casn-chapter}, we describe how an operation \texttt{Op} should be executed in a node \texttt{v}.

Since now we force processes to help each other, operation \texttt{Op}, initiated by process \texttt{P}, in any node \texttt{v} can be executed by some other helper process. % (e.g., while \texttt{P} is yielded by the underlying OS scheduler).
Thus, we need to provide a mechanism for the process \texttt{P} by which it distinguishes between the two following situations:

\begin{itemize}
    \item Operation \texttt{Op} has not yet been executed in node \texttt{v}. Thus, the descriptor of \texttt{Op} is still located somewhere in \texttt{v} queue. In that case, \texttt{P} needs to continue executing operations from the head of \texttt{v} queue in node \texttt{v}.
    
    \item Operation \texttt{Op} has already been executed in node \texttt{v}. In that case, \texttt{P} can proceed to execute \texttt{Op} in lower nodes of \texttt{v}'s subtree.
\end{itemize}

We use timestamps to distinguish between these two situations. We describe that usage of timestamps with formulating and proving \emph{timestamps increasing property}. 

\begin{theorem}
In each queue, operation timestamps form a monotonically increasing sequence. More formally, if at any moment we traverse any queue \texttt{Q} from the head to the tail and obtain $\texttt{t}_{\texttt{1}}, \texttt{t}_{\texttt{2}}, \ldots \texttt{t}_{\texttt{n}}$~--- a sequence of timestamps of descriptors, located in \texttt{Q}, then $\texttt{t}_{\texttt{1}} < \texttt{t}_{\texttt{2}} < \ldots < \texttt{t}_{\texttt{n}}$ will hold.
\end{theorem}

We prove that theorem in Appendix~\ref{timestamp-monotony-proof-section}.

As follows from that property, the initiator process \texttt{P} can easily learn, whether its operation \texttt{Op} has been executed in node \texttt{v} by using the simple algorithm: 

\begin{itemize}
    \item if the queue is empty~--- we conclude that \texttt{Op} has been executed in \texttt{v};
    \item if the queue is not empty, we compare the timestamp of the descriptor in the head of \texttt{v} queue with the timestamp of \texttt{Op}: if \texttt{v.Queue.Head.Timestamp $>$ Op.Timestamp}, we conclude that \texttt{Op} has been executed in \texttt{v}, otherwise, we conclude that \texttt{Op} has not been executed in \texttt{v} yet.
\end{itemize}

Therefore, we can implement the algorithm of executing all operations from \texttt{v}'s queue up to \texttt{Op.Timestamp} (Listing~\ref{execute-until-timestamp-listing}):

\renewcommand{\lstlistingname}{Listing}
\begin{lstlisting}[caption={The algorithm to execute all operations, up to the specified timestamp \texttt{Op.Timestamp}, from \texttt{v} queue},label={execute-until-timestamp-listing},escapeinside={(*}{*)}, captionpos=b]
fun execute_until_timestamp(Op, v):
    while true:
        // obtains the first descriptor in FIFO order
        head_descriptor := v.Queue.peek()
        if head_descriptor = nil:
            return
        if head_descriptor.Timestamp > Op.Timestamp:
            return
        // execute_in_node changes states of v children
        // pushes head_descriptor to child queues, 
        // removes head_descriptor from v queue 
        execute_in_node(head_descriptor, v)
\end{lstlisting}

Suppose the initiator process \texttt{P} is traversing the tree to execute operation \texttt{Op} and \texttt{P} just finished executing \texttt{Op} in node \texttt{v}. How can \texttt{P} choose the next node in the traversal? It is not necessary to always continue the traversal in one of \texttt{v}'s children, since \texttt{Op} can be now finished in \texttt{v} subtree by other helper processes. To address this issue, in each operation descriptor we store a queue with nodes \texttt{Op.Traverse}~--- the queue of nodes that must be visited during the execution of \texttt{Op}. The \texttt{Traverse} queue is maintained and used in the following way:

\begin{itemize}
    \item When any process (no difference initiator or helper) starts executing \texttt{Op} in node \texttt{v}, it adds to the tail of \texttt{Op.Traverse} all children of node \texttt{v} in which the execution should continue;

    \item When the initiator process finishes the procedure \texttt{execute\_until\_timestamp(Op.Timestamp, v)}, it removes \texttt{v} from the head of \texttt{Op.Traverse} queue. Note, that only the initiator process can remove nodes from \texttt{Op.Traverse} queue;

    \item After the initiator process has removed the current node \texttt{v} from the head of \texttt{Op.Traverse}, it checks \texttt{Op.Traverse}: if it is empty, the operation is completed and the initiator returns the query result to the caller; otherwise (if \texttt{Op.Traverse} is not empty), the initiator continues the traverse by taking the next node from the head of \texttt{Op.Traverse}.
\end{itemize}

Note, that this queue maintenance scheme allows a node \texttt{v} to be inserted into \texttt{Op.Traverse} multiple times, since multiple helper processes may be executing \texttt{Op} in \texttt{v} parent in parallel. However, as will be explained in Section~\ref{no-casn-chapter} \texttt{v}'s state will still be modified exactly once, no matter how many times it is processed. The traverse algorithm can be implemented as in Listing~\ref{traverse-listing}.

%\ik{Draw some picture about multiple v in queue here?}

\renewcommand{\lstlistingname}{Listing}
\begin{lstlisting}[caption={The algorithm for traversing the tree},label={traverse-listing},escapeinside={(*}{*)}, captionpos=b]
fun execute_operation(op):
    Tree.Root.Queue.push_acquire_timestamp(op)
    op.Traverse = {Tree.Root}
    while true:
        v := op.Traverse.peek()
        if v = nil: // op is finished
            return
        execute_until_timestamp(op.Timestamp, v)
        op.Traverse.pop()
\end{lstlisting}

Now we have to design a method, that will allow the initiator process to learn the operation result when the operation is completed. The problem here is that the operation result might consist of multiple parts (e.g., \texttt{count} result consists of a sum of multiple subtree sizes), and these parts (e.g., subtree sizes) may be computed by different processes, since force processes to help each other. 

To allow operation result to be assembled from these parts, in each operation descriptor we store a concurrent map \texttt{Op.Processed}, filled with nodes, in which the execution of \texttt{Op} has been finished. The size of this map is expected to be small for aggregate range queries (e.g., $O(\log N)$), so, we can implement them in any way we want: a wait-free queue that stores all the required nodes (maybe multiple times, which we filter out at the end of the operation) or with a Wait-free Universal Construction~\cite{herlihy1991wait}, and, finally, we can use a wait-free map.

The \texttt{Op.Processed} uses tree nodes as its keys. To allow this, we augment each tree node \texttt{v} with an identifier, stored in the \texttt{v.Id} field. Each node receives its identifier at the creation moment and the node identifier does not change throughout the node lifetime. The node identifiers must be unique. We can achieve that property using UUID~\cite{uuid-wiki} generation procedure or by incrementing \texttt{fetch-and-add}~\cite{fetch-and-add} counter.

Values of the \texttt{Op.Processed} map store parts of the result: for example, for the \texttt{count} query we store in the \texttt{Op.Processed} the node identifiers with the sizes of their subtrees that should be added to the result of the query.

Before removing \texttt{Op} descriptor from the head of \texttt{v}'s queue we try to add \texttt{v.Id} along with a value \texttt{x}, corresponding to the part of the answer for the node \texttt{v}, into the \texttt{Op.Processed} map. If key \texttt{v.Id} already exists in the \texttt{Processed} map, we left the \texttt{Op.Processed} map unmodified, without changing the value, associated with \texttt{v.Id}.

We never modify the value, associated with node \texttt{v}, since stalled processes can calculate the value incorrectly. Indeed, consider the following scenario:

\begin{nenum}
    \item Descriptor \texttt{D}, corresponding to a \texttt{count} operation with timestamp \texttt{42}, is located at the head of \texttt{v}'s queue;

    \item Process \texttt{P} reads \texttt{D} from the head of \texttt{v}'s queue;
    
    \item Process \texttt{P} is suspended by the OS;
    
    \item Process \texttt{R} reads \texttt{D} from the head of \texttt{v}'s queue;
    
    \item Process \texttt{R} determines that the size of \texttt{v}'s left subtree should be added to the result;
    
    \item Process \texttt{R} reads the size of \texttt{v}'s left subtree (say, it equals to 5) and adds key-value pair \texttt{$\langle$ v.Id, 5 $\rangle$} to the \texttt{Processed} map;
    
    \item A new key is inserted to \texttt{v} left subtree by \texttt{insert} operation with timestamp \texttt{43}, making \texttt{v} left subtree size equal to 6;
    
    \item Process \texttt{P} is resumed by the OS;
    
    \item Process \texttt{P} reads the size of \texttt{v}'s left subtree (now it equals to 6) and tries to add key-value pair \texttt{$\langle$ v.Id, 6 $\rangle$} to the \texttt{Processed} map.
\end{nenum}

On step (9) we should not modify the value, corresponding to the node \texttt{v}, since the value \texttt{6} reflects the modification, performed by the \texttt{insert} operation with timestamp \texttt{43}. The \texttt{count} operation has timestamp \texttt{42}, thus, the \texttt{count} result should not include the key, inserted by \texttt{insert} operation with timestamp \texttt{43}.

When the operation execution is finished (i.e., \texttt{Op.Traverse} is empty) we traverse the \texttt{Processed} map, forming the query result from partial results associated with visited nodes. Note, that it is safe to traverse the \texttt{Processed} map~--- indeed, now the \texttt{Processed} map cannot be modified concurrently, since the query execution is finished.

\subsection{Detailed description of an execution in a node}
\label{no-casn-chapter}

% The algorithm to execute operation \texttt{Op} in node \texttt{v} consists of the following steps:

% \begin{nenum}
%     \item Determine the set of children \texttt{C}, in which execution of \texttt{Op} should continue.
    
%     \item Traverse the set \texttt{C}. For each child \texttt{c} from \texttt{C}:
%         \begin{nenum}
%             \item Atomically read the state of \texttt{c}.
            
%             \item If the state of \texttt{c} has not been modified by \texttt{Op} yet, modify it. We explain how to do it below.
            
%             \item Insert \texttt{Op} descriptor to the queue of \texttt{c} if it has not been yet inserted.
%         \end{nenum}
    
%     \item If \texttt{Op} descriptor has not been yet removed from \texttt{v} queue, remove it.
%     %We explain in Section~\ref{pop-if-section} how to do it.
% \end{nenum}

In Section~\ref{operation-execution-chapter}, we explained how the execution of the operation works in general. Now, we go into details of the execution in the node.

The process of executing an operation \texttt{Op} in a node \texttt{v} consists of the following actions:
\begin{itemize}
    \item Determine the set of child nodes \texttt{C}, in which \texttt{Op} execution should continue. 
    
    \item For each child \texttt{c} from the set \texttt{C}:
    \begin{nenum}
        \item Insert \texttt{c} into \texttt{Op.Traverse} queue;
        
        \item Modify the state of \texttt{c} (e.g., a size of \texttt{c}'s subtree), if necessary;
        
        \item Insert \texttt{Op} descriptor to the operations queue of \texttt{c}, thus allowing \texttt{Op} to continue its execution at lower levels of the tree.
    \end{nenum}

    \item Try to add \texttt{v.Id} along with a value \texttt{x}, corresponding to the part of the answer for the node \texttt{v}, into \texttt{Op.Processed} map.
    %If \texttt{Op.Processed} map already contains entry with key \texttt{v.Id}, leave \texttt{Op.Processed} unmodified.
    
    \item Try to remove \texttt{Op} descriptor from the head of \texttt{v}'s queue if it is still there.
\end{itemize}

The removal of \texttt{Op} descriptor from the head of \texttt{v}'s queue should be done after the insertion of \texttt{Op} descriptor to child queues and modification of child states are finished. Otherwise, the execution of later operations in \texttt{v} may start before the execution of \texttt{Op} in \texttt{v} is finished, which may break the main invariant (Section~\ref{main-invariant-chapter}). 
Inserting the descriptor to child queues, modifying child states, and removing the descriptor from the parent queue should happen exactly once, no matter how many processes are working on the descriptor concurrently.

Exactly-once insertion to and removal from queues is handled by our implementation of concurrent queues (see Section~\ref{queue-section}). Queues provide two procedures:

\begin{itemize}
    \item \texttt{push\_if} inserts the descriptor to the tail of the queue only if it has not been inserted yet, otherwise, the queue is left unmodified.
    %The implementation of this procedure is discussed in Section~\ref{push-if-section}.

    \item \texttt{pop\_if} removes the descriptor from the head of the queue only if it has not been removed yet, otherwise, the queue is left unmodified.
    %The implementation of this procedure is discussed in Section~\ref{pop-if-section}.
\end{itemize}

The main problem in the execution of an operation \texttt{Op} in a node \texttt{v} is the proper work with the children states: we should be able to work with each state atomically and we should modify each state exactly once, no matter how many processes are executing \texttt{Op} in \texttt{v}.

The atomicity problem comes from the fact that the state may consist of multiple fields. To solve this problem, we do not store the state directly inside the node~--- instead we store the immutable state in the heap and the node stores the pointer \texttt{S\_Ptr} to it.

The state, located in the heap, is considered immutable and is never modified. To modify the node state, we simply do the following:

\begin{nenum}
    \item create the structure, corresponding to the modified state, with an arbitrary set of fields changed;
    \item place the modified state somewhere in the heap;
    \item change the \texttt{node.S\_Ptr}, so that it points to the new state.
\end{nenum}

To read the state atomically, we simply read the \texttt{S\_Ptr} register. After that, we can safely access any field from the state structure, pointed at by the fetched pointer, without worrying that the state structure is being modified concurrently by another process. Since the structure is immutable, it can never be modified by another process.

Now, we return to the second problem of modifying the state exactly once. In the state we store one additional field: \texttt{Ts\_Mod}~--- timestamp of the operation, that was the last to modify the state. Thus, if the operation \texttt{Op} is willing to modify the state of node \texttt{v}, we should first read the current \texttt{v}'s state and acquire the last modification timestamp.

\begin{itemize}
    \item If \texttt{Ts\_Mod $\geq$ Op.Timestamp} we conclude that \texttt{v}'s state has been already modified by \texttt{Op}. In that case, we simply do not try to modify \texttt{v}'s state according to \texttt{Op} anymore.
    
    \item Otherwise, we create a new state (with \texttt{Ts\_Mod = Op.Timestamp}) and try to change the state pointer using \texttt{CAS(\&v.S\_Ptr, cur\_state, new\_state)}. We then go to the next step, no matter what was the \texttt{CAS} result. If the \texttt{CAS} returned \texttt{true}~--- we have successfully modified the state, otherwise (if the \texttt{CAS} returned \texttt{false}), some other process has already modified the state according to \texttt{Op}.
\end{itemize}

Thus, the state is modified with each executed operation exactly once. Indeed, even if some stalled process will try to modify node \texttt{v} with an already applied operation \texttt{Op} the node state will not be changed, since the last modification timestamp is greater than or equal to \texttt{Op.Timestamp}. 
Therefore, the algorithm can be implemented in the following way (Listing~\ref{op-process-node-no-casn-listing}):

\renewcommand{\lstlistingname}{Listing}
\begin{lstlisting}[caption={Algorithm for executing operation \texttt{op} in node \texttt{v}},label={op-process-node-no-casn-listing},escapeinside={(*}{*)}, captionpos=b]
fun execute_in_node(op, v):
    C := /* set of v children in which 
        execution of op should continue */
    for c in C:
        cur_state := v.State_Ptr
        op.Traverse.push(c)
        if cur_state.Ts_Mod < op.Timestamp:
            new_state := op.get_modified_state(cur_state)
            new_state.Ts_Mod := op.Timestamp
            CAS(&v.State_Ptr, cur_state, new_state)
        c.Queue.push_if(op)
    node_key := v.Id
    node_value := /* part of the result 
        corresponding to v */
    op.try_insert(node_key, node_value)
    v.Queue.pop_if(op)
\end{lstlisting}

\subsection{Implementation of an operations queue}
\label{queue-section}

\textbf{Queue structure.}
For our purpose, we can use any practical queue algorithms as a basis for our descriptors queues, e.g., \texttt{fetch-and-add} queue~\cite{yang2016wait} or practical wait-free queue~\cite{kogan2011wait}: the final implementation remains almost the same. However, for simplicity of the presentation, we use Michael-Scott queue. This queue is lock-free which makes the whole algorithm lock-free. But if we make the root queue to be wait-free~--- all other queues based on Michael-Scott queue will automatically have the same progress guarantee due to the way how we work with the descriptors. For more information about the wait-freedom see Section~\ref{wait-free-section}.

In each node of the queue we store the descriptor in field \texttt{Data} and the pointer to the next node in field \texttt{Next}.
Also, we have two pointers: \texttt{Tail}, that points to the last node of the queue, and \texttt{Head}, that points to the node \emph{before} the first node of the queue. Note that the node at \texttt{Head} pointer does not store any data, residing in the queue. This node is considered dummy and only the node at \texttt{Head.Next} pointer contains the first real descriptor in the queue.

\textbf{Queue in the root.}

As discussed in Section~\ref{main-invariant-chapter}, the operation queue in the root node should provide timestamp allocation mechanism, with the following guarantees: if the descriptor of operation \texttt{A} was added to the root queue before the descriptor of the operation \texttt{B}, then \texttt{timestamp(A) < timestamp(B)} should hold. 

% Note, that the descriptor becomes visible to all the system processes at the moment it is added to the root queue, and, as described in Section~\ref{operation-execution-chapter}, the system processes examine timestamps of all descriptors in order to execute their operations. Thus, the timestamp should be written to the \texttt{descriptor.Timestamp} field before the descriptor is added to the root queue.

As stated above, we can use a slight modification of Michael-Scott queue~\cite{michael1996simple} to implement the timestamp allocation mechanism for the root queue.
Each time we need to add a new descriptor to the root queue, we 1)~create a new node with the descriptor; 2)~take the timestamp of the tail; 3)~set the new timestamp in our descriptor as the incremented timestamp of the tail; 4)~try to move the queue tail to the new node using CAS; 5) if the CAS is successful we stop, otherwise, we repeat from step (2).

% The algorithm can be implemented the following way (Listing~\ref{push-acquire-timestamp-listing}):

% \renewcommand{\lstlistingname}{Listing}
% \begin{lstlisting}[caption={Implementation of the \texttt{push} procedure with acquiring operation timestamp},label={push-acquire-timestamp-listing},escapeinside={(*}{*)}, captionpos=b]
% /* 
% Executed by the initiator process at the 
% beginning of the operation execution 
% */
% fun push_acquire_timestamp(Root_Queue, descriptor):
%     new_node := new QueueNode(Data = descriptor, Next = nil)
%     while true:
%         cur_tail := Root_Queue.Tail
%         max_timestamp := cur_tail.Data.Timestamp
%         descriptor.Timestamp (*$\leftarrow$*) max_timestamp + 1
%         if CAS(&cur_tail.Next, nil, new_node):
%             CAS(&Root_Queue.Tail, cur_tail, new_node)
%             return
%         else:
%             other_process_tail := cur_tail.Next
%             CAS(&Root_Queue.Tail, cur_tail, other_process_tail)
%             /* Retry the whole operation from the very beginning */
% \end{lstlisting}

In Section~\ref{wait-free-section}, we show how to implement such queue in a wait-free manner. %, for example, by using the Universal Construction~\cite{herlihy1991wait}.
%For more information see Section~\ref{wait-free-section}.

\textbf{\texttt{push\_if} implementation.}
As discussed in Section~\ref{no-casn-chapter}, non-root queues should provide \texttt{push\_if} operation that inserts a descriptor into the queue if it was not inserted yet (otherwise, the queue should be left unmodified). The procedure is based on the Michael-Scott queue insertion algorithm~\cite{michael1996simple}: we check the timestamp of the tail, if it is higher then the descriptor has been inserted and we leave the queue unmodified, otherwise, we try to move the queue tail to the new node using CAS.

% Therefore, the algorithm can be implemented the following way (Listing~\ref{push-if-listing}):

% \renewcommand{\lstlistingname}{Listing}
% \begin{lstlisting}[caption={Implementation of the \texttt{push\_if} procedure },label={push-if-listing},escapeinside={(*}{*)}, captionpos=b]
% fun push_if(Non_Root_Queue, descriptor):
%     new_node := new QueueNode(Data = descriptor, Next = nil)
%     while true:
%         cur_tail := Non_Root_Queue.Tail
%         if cur_tail.Data.Timestamp (*$\geq$*) descriptor.Timestamp:
%             return
%         elif CAS(&cur_tail.Next, nil, new_node):
%             CAS(&Non_Root_Queue.Tail, cur_tail, new_node)
%             return
%         else:
%             other_process_tail := cur_tail.Next
%             CAS(&Non_Root_Queue.Tail, cur_tail, other_process_tail)
%             /* Retry the whole operation from the very beginning */
% \end{lstlisting}

\textbf{\texttt{pop\_if} implementation.}
As discussed in Section~\ref{no-casn-chapter}, the operation queue in any node should provide \texttt{pop\_if} operation, that tries to remove descriptor with the specified timestamp \texttt{TS} from the head of the queue. If descriptor \texttt{D} with timestamp \texttt{TS} is still located at the head of the queue, it is removed. Otherwise, the queue is left unmodified~--- in this case, we assume that \texttt{D} was removed by some other process. We assume that at some moment \texttt{D} was located at the head of the queue (it may still be located at the head of the queue or it may be already removed), i.e., we never try to remove a descriptor from the middle of the queue. We can do this using Michael-Scott queue~\cite{michael1996simple}.

% Therefore, \texttt{pop\_if} can be implemented in the following way (Listing~\ref{pop-if-listing}):

% \renewcommand{\lstlistingname}{Listing}
% \begin{lstlisting}[caption={Implementation of the \texttt{pop\_if} procedure },label={pop-if-listing},escapeinside={(*}{*)}, captionpos=b]
% fun pop_if(Queue, timestamp):
%     while true:
%         cur_head := Queue.Head
%         cur_tail := Queue.Tail
%         if cur_head = cur_tail:
%             next_tail := cur_tail.Next
%             if next_tail = nil:
%                 return
%             else:
%                 CAS(&Queue.Tail, cur_tail, next_tail)
%                 /* Retry the operation from the very beginning */
%         else:
%             next_head := cur_head.Next
%             first_timestamp := next_head.Data.Timestamp
%             if first_timestamp = timestamp:
%                 CAS(&Queue.Head, cur_head, next_head)
%             return
% \end{lstlisting}

\subsection{Balancing strategy}
\label{balancing-section}

Until now, we considered unbalanced trees which may have $height \in \Omega(\log N)$. Since most of the queries (e.g., \texttt{insert}, \texttt{remove}, \texttt{contains} ,and \texttt{count}) are executed on a tree in $\Theta(height)$ time, using unbalanced trees may result in these queries being executed in non-optimal $\omega(\log N)$ time. Therefore, we must design an algorithm to keep the tree balanced. One possible balancing strategy is based on a subtree rebuilding and is similar to the balancing strategy proposed in~\cite{brown2020non,prokopec2020analysis,aksenov2023parallel,mehlhorn1993dynamic}. The idea of this approach can be formulated the following way: when the number of modifications in a particular subtree exceeds a threshold, we rebuild that subtree making it perfectly balanced.% (Fig~\ref{rebuild-idea-pic}).

% \begin{figure}[H]
%   \centering
%   \caption{Tree balancing via subtree rebuilding: when the number of modifications, applied to a subtree, exceeds a threshold, we rebuild the whole subtree}
%   \label{rebuild-idea-pic}
%   \includegraphics[width=0.5\linewidth]{pics/rebuild-idea.png}
% \end{figure}

For each tree node we maintain \texttt{Mod\_Cnt} in the node state~--- the number of modifications in the subtree of this node. Moreover, for each node we store an immutable number \texttt{Init\_Sz}~--- the initial size of its subtree, i.e., the number of data items in that node subtree at the moment of node creation (node can be created when a new data item is inserted to the tree or when the subtree, where the node is located, is rebuilt). We rebuild the node subtree when \texttt{Mod\_Cnt > K $\cdot$ Init\_Sz}, where \texttt{K} is a predefined constant. This approach makes the rebuilding to take $O(1)$ amortized time and, thus, the rebuilding does not affect amortized total cost (according e.g., to~\cite{mehlhorn1993dynamic}).

We check whether the subtree of \texttt{v} needs rebuilding (and perform the rebuilding itself) before inserting an operation descriptor to \texttt{v}'s queue and changing \texttt{v}'s state. Therefore, we can perform \texttt{v}'s subtree rebuilding only during execution of some operation in \texttt{v}'s parent. 

Consider node \texttt{v}, its parent \texttt{pv} and operation \texttt{Op}, that is being executed in \texttt{pv} and that should continue its execution in \texttt{v}'s subtree (and, therefore, its descriptor should be inserted to \texttt{v}'s queue). Before inserting \texttt{Op} to \texttt{v}'s queue and changing \texttt{v} state, we check whether \texttt{Mod\_Cnt} in \texttt{v} will exceed the threshold after applying \texttt{Op} to \texttt{v}'s subtree: if so, \texttt{v} subtree must be rebuilt.

\begin{figure}[H]
  \centering
  \caption{The subtree that needs rebuilding may contain descriptors of unfinished operations}
  \label{rebuild-unfinished-rebuild-pic}
  \includegraphics[width=\linewidth]{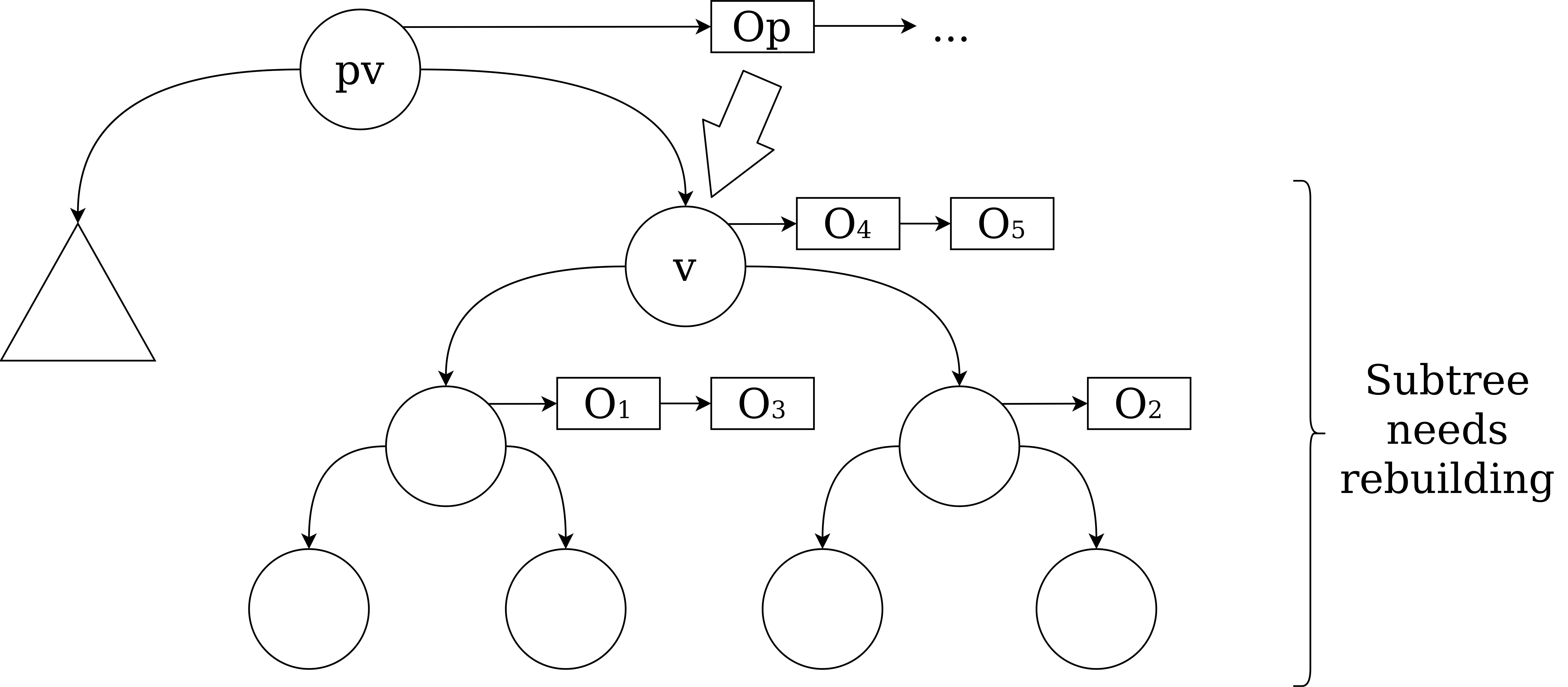}
\end{figure}

Note, that the subtree of \texttt{v} can contain unfinished operations: their descriptors still reside in the queues in that subtree (Fig.~\ref{rebuild-unfinished-rebuild-pic}).

As the first step, we should finish all these unfinished operations before rebuilding the subtree. To do so, we traverse the subtree and in each node \texttt{u $\in$ subtree(v)} execute all operations, residing in \texttt{u} queue.
After that, we again traverse the subtree of \texttt{v}, that no longer contains unfinished operations, and collect all the stored data items (e.g., keys or key-value pairs). Then, we build an ideally balanced subtree, containing all these data items.

Each node of the new subtree should be initialized with \texttt{Mod\_Cnt = 0} and contain correct \texttt{Init\_Sz}. We should set \texttt{Ts\_Mod} of each node in the rebuilt subtree so that \texttt{Op} and all later operations (with \texttt{timestamp $\geq$ Op.Timestamp}) can still modify the new subtree, but all the preceding operations (with \texttt{timestamp < Op.Timestamp}) cannot. Thus, we set \texttt{Ts\_Mod = Op.Timestamp - 1}.

After that, we take \texttt{nv}~--- the root of the new subtree and try to modify the pointer that pointed at \texttt{v}, so that it starts to point at \texttt{nv}. For example, if \texttt{v} was the left child of \texttt{pv}, we execute \texttt{CAS(\&pv.Left, v, nv)}; if \texttt{v} was the right child of \texttt{pv}, we execute \texttt{CAS(\&pv.Right, v, nv)}. 
If the \texttt{CAS} returned \texttt{true} we conclude that we have successfully finished the rebuilding; if the \texttt{CAS} returned \texttt{false} we conclude that some other process has completed the rebuilding before us. In either case we resume the execution of \texttt{Op} in \texttt{pv}: we read \texttt{nv}~--- new root of the subtree, modify \texttt{nv}'s state, insert \texttt{Op} descriptor to \texttt{nv}'s queue (here we re-read root of the subtree because \texttt{nv} can be root of the subtree build not by our process, but by some another helder process) and remove \texttt{Op} descriptor from \texttt{pv} queue.

% Indeed, if \texttt{CAS} returned \texttt{true}, we conclude that we have successfully completed the rebuilding and can proceed with the execution of \texttt{Op} in \texttt{pv}. Otherwise, if \texttt{CAS} returned \texttt{false}, we conclude that the rebuilding was completed by some other process and it already modified the necessary child pointer. Yet again, in that case we proceed with the execution of \texttt{Op} in \texttt{pv}.

% Note that new descriptors cannot appear in \texttt{v} subtree  until the rebuild is completed, since new descriptors cannot be inserted to \texttt{v} queue until the execution of \texttt{Op} in \texttt{pv} is finished, according to the main invariant (Fig.~\ref{rebuild-wait-pic}). Since the execution of \texttt{Op} in \texttt{pv} includes rebuilding \texttt{v} subtree, all the later descriptors are inserted to the rebuilt subtree.

% \begin{figure}[H]
%   \centering
%   \caption{New descriptors cannot appear in \texttt{v} subtree during the rebuilding procedure}
%   \label{rebuild-wait-pic}
%   \includegraphics[width=\linewidth]{pics/rebuild-wait.png}
% \end{figure}

\subsection{Wait-freedom}
\label{wait-free-section}

We now prove that our solution can be implemented efficiently with wait-free progress guarantee.
We recall that \emph{wait-freedom}~\cite{herlihy1991wait} is a progress guarantee that requires all non-suspended processes to finish their execution within a bounded number of steps.

\begin{theorem}
Each operation \texttt{Op} in our solution finishes within a bounded number of steps.
\end{theorem}

To prove that theorem we recall that the execution of operation \texttt{Op} consists of:

\begin{enumerate}
    \item Inserting \texttt{Op} descriptor into the root queue;
    \item Propagating \texttt{Op} descriptor downwards, from the root to the appropriate lower nodes;
    \item Executing \texttt{Op} in each node \texttt{v} on the target path.
\end{enumerate}

Now, we prove that each of these stages finishes within a bounded number of steps.
    
\begin{lemma}
    The insertion of a descriptor into the root queue finishes within a bounded number of steps.
\end{lemma}

\begin{proof}
    Our queue implementation, described in Section~\ref{queue-section} is lock-free, but not wait-free, since it is just a version of Michael and Scott queue~\cite{michael1996simple}.

    The simplest approach is to implement the wait-free root queue using the well-known Wait-free Universal Construction~\cite{herlihy1991wait}, with no implementation caveats.

    However, this approach has a very huge overhead. We hope that some practical wait-free queue (e.g.,~\cite{yang2016wait,kogan2011wait}) can emulate our root queue and its timestamps distribution.
    Unfortunately, a wait-free queue from~\cite{yang2016wait} can support the increasing timestamps using cell identifiers for that, but do not allow a simple wait-free peek function, that reads the head of the queue but does not remove it~--- this functionality is crucial for our queue in \texttt{pop\_if}.
    Luckily for us the wait-free queue from~\cite{kogan2011wait} supports wait-free peek function and supports non-decreasing timestamps (or \texttt{epochs} in the paper). We can make them strongly increasing using a \texttt{fetch-and-add} register.
    %For example, in~\cite{yang2016wait} each value is stored in some cell which identifier could be seen as a timestamp. Thus, we just need to set the timestamp of the descriptor to a cell identifier during the \texttt{enqueue} operation right before the operation writes our descriptor into the cell: it can be done by a small update of \texttt{enq\_fast} and \texttt{enq\_commit} functions. This way we have descriptors with increasing timestamps in the corresponding order in this queue.

    %Now, we present the wait-free algorithm for the queue with our requirements with small overhead.
    %At first, note that our main target is to give the timestamps to the descriptors. Then, we can also use the queue with timestamps in the root as explained in Section~\ref{queue-section}.
    To distribute the timestamps, we need a version variable and an array of size $P$ that contains the current descriptors. Each descriptor has an empty timestamp variable at the initialization. When performing an operation, process $\pi$ creates a new descriptor and puts it into the corresponding cell. Then, it gets new version from the version variable using \texttt{fetch-and-add} and tries to CAS the current empty timestamp in its descriptor to the obtained version. Not depending on the result of CAS, the descriptor of $\pi$ has a timestamp. Then, $\pi$ traverses the array of descriptors and replaces empty timestamps by a newly fetched version. Also, $\pi$ saves the descriptors with the timestamp smaller than the one in its descriptor. Finally, the process tries to enqueue into the root queue all these descriptors in the sorted order of their timestamps. Thus, the algorithm works in $O(P \log P)$ time.
\end{proof}

\begin{lemma}
\label{lem:onenode}
In each tree node \texttt{v} on the \texttt{Op} traversal path executing \texttt{Op} in \texttt{v} finishes in a finite number of steps.
\end{lemma}

\begin{proof}
    Consider an operation queue at node \texttt{v} (Fig.~\ref{operation-execution-waitfree-pic}). Here some operations ($X_1 \ldots X_K$) should be executed before \texttt{Op}, while all other operations ($Y_1 \ldots$) will be executed only after execution of \texttt{Op} in \texttt{v} is fully completed. Thus:
    
    \begin{figure}[H]
      \centering
      \caption{Operation queue structure at node \texttt{v}}
      \label{operation-execution-waitfree-pic}
      \includegraphics[width=\linewidth]{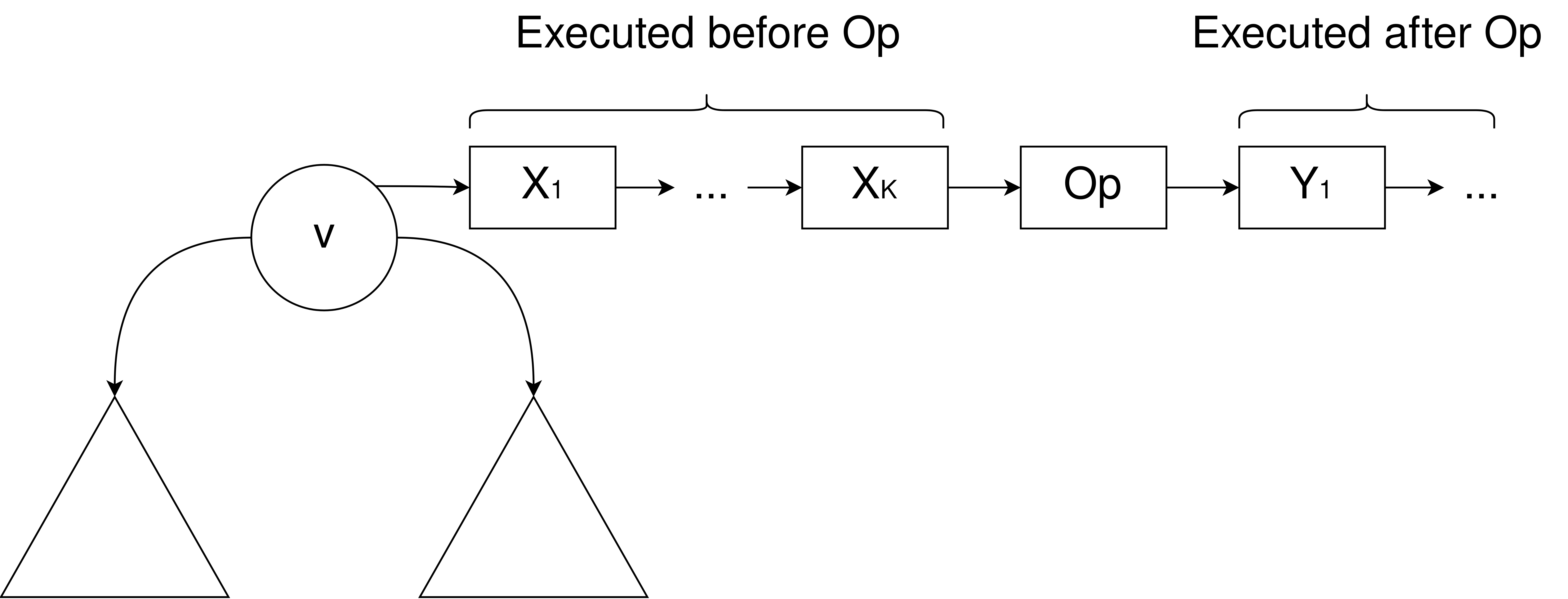}
    \end{figure}

    \begin{itemize}
        \item We help to complete only a finite number of operations in a node \texttt{v}, since there cannot be more than $\vert P \vert$ operations in the queue of \texttt{v} before \texttt{Op} (where $P$ is the set of the processes executing operations);

        \item Each operations $X_i$ takes a finite number of steps to complete its execution in a node \texttt{v} (see Section~\ref{no-casn-chapter} for the list of those steps). Note, that in the process of execution operation \texttt{Op} in node \texttt{v} we never retry any operation (in contrast to lock-free algorithms, e.g., in \cite{michael1996simple}): for example if the insertion of \texttt{Op} descriptor to child node \texttt{cv} fails, we conclude that \texttt{Op} descriptor has been inserted to \texttt{cv} by another helper process and merely continue the execution of \texttt{Op} in \texttt{v};
    \end{itemize} 
    
    Therefore, executing \texttt{Op} in \texttt{v} finishes in a constant number of steps.
\end{proof}

\begin{lemma}
\label{lem:downwards}
Propagating the descriptor downwards, from the root to the appropriate lower nodes finishes within a bounded number of steps
\end{lemma}

\begin{proof}
Consider some operation $Op_2$ such that \texttt{$Op_2$.Timestamp > Op.Timestamp}. If both $Op_2$ and \texttt{Op} are willing to change the very same tree node \texttt{v}, $Op_2$ under any conditions will do it after \texttt{Op}, since the operations are executed in a strict timestamp order (see Section~\ref{main-invariant-chapter} for details). Thus, $Op_2$ cannot somehow change the structure of the tree to disrupt \texttt{Op}'s traversal. Therefore, \texttt{Op} will finish its traversal in a constant amount of steps, since later operations cannot interfere in \texttt{Op} traversal. Since none of the later operations can overcome \texttt{Op}, we note the following:

\begin{itemize}
    \item At the moment when \texttt{Op} begins execution the size of tree is $N$ and no more than $\vert P \vert$ concurrent processes are inserting new nodes in the tree. Thus at the \texttt{Op.Timestamp} moment the size of the tree will no exceed $O(N + \vert P \vert$), which is definitely a finite number;

    \item By Lemma~\ref{lem:onenode} operations takes a finite number of steps to execute in a node. 
\end{itemize}

Thus, the operation takes a finite number of steps to finish its traversal.
\end{proof}

Note, that our rebuilding procedure does not fail the wait-freedom guarantee in the proof above since each rebuilding finishes in a bounded number of steps. 
\begin{lemma}
\label{lem:rebuilding}
The rebuilding procedure finishes in a bounded number of steps
\end{lemma}

\begin{proof}
Indeed, the rebuilding procedure of a subtree \texttt{vs} consists of the following steps:

\begin{itemize}
    \item Traverse the subtree \texttt{vs}, collecting all unfinished operations;
    \item Help to complete all these unfinished operations;
    \item Collect all keys from \texttt{vs};
    \item Build an ideal tree from collected keys.
\end{itemize}

Note, that only the operations that started before \texttt{Op} can be unfinished in \texttt{vs} (Fig~\ref{rebuild-unfinished-pic}), since we execute operations in the timestamp order. 

\begin{figure}[H]
  \centering
  \caption{Unfinished operation $O_1, O_2, \ldots O_5$ have timestamp lower than \texttt{Op.Timestamp}}
  \label{rebuild-unfinished-pic}
  \includegraphics[width=\linewidth]{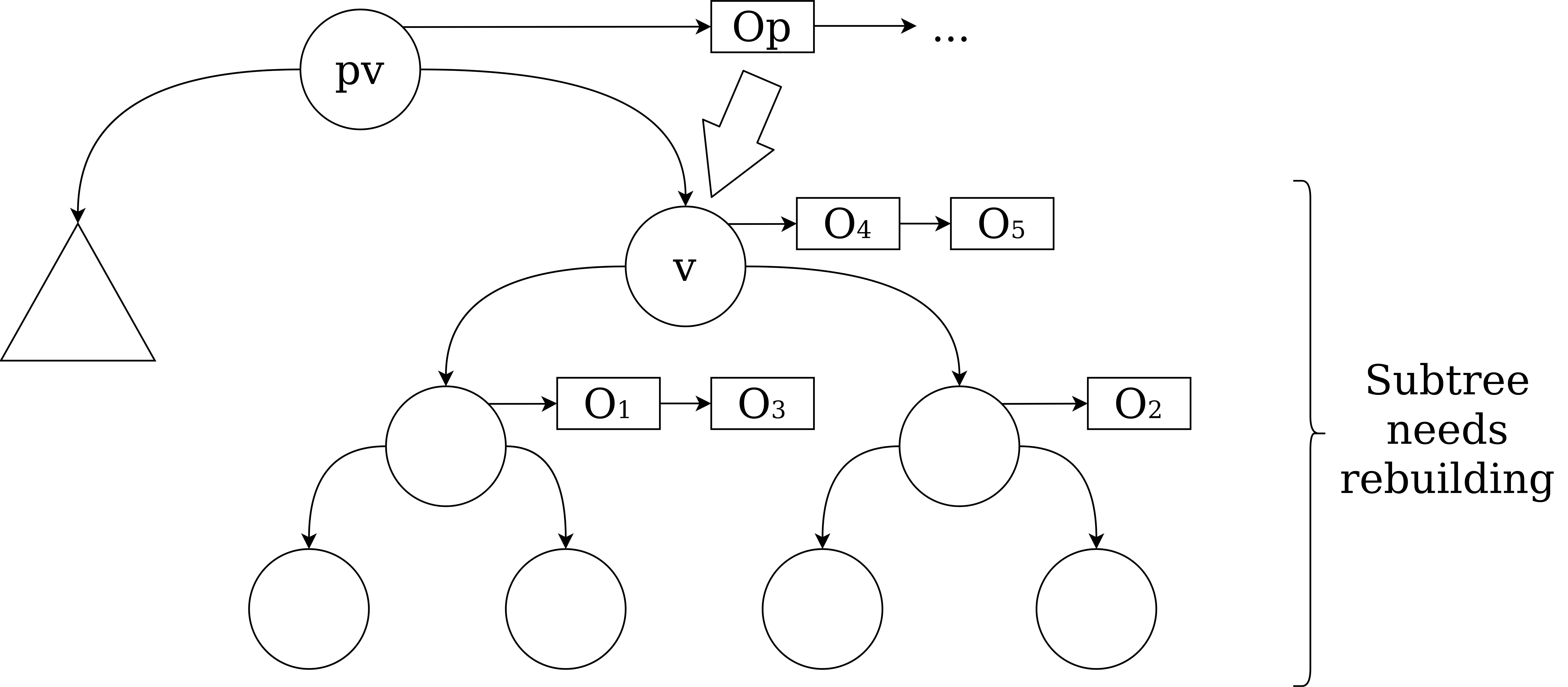}
\end{figure}

Therefore: 1)~there is a finite set of unfinished operations in \texttt{vs}; 2)~a completing of each unfinished operations takes a finite number of steps by Lemma~\ref{lem:downwards};
3)~\texttt{vs} has a finite size, thus, the collecting all keys from \texttt{vs} and the construction of a new ideal subtree also takes a finite amount of steps.
Thus, the rebuilding completes in a finite amount of steps.
\end{proof}

\subsection{Time cost analysis}

We now estimate the time it takes to execute an operation in our solution.

\begin{theorem}
    The amortized cost of \texttt{insert}, \texttt{remove}, \texttt{contains} or \texttt{count} operation on our concurrent binary search tree with rebuilding is $O((\log N + \vert P \vert) \cdot \vert P \vert)$
\end{theorem}

\begin{proof}
    Suppose $N$ is the size of the tree when the operation \texttt{Op} starts its execution. In a sequential setting each of these operations takes $O(\log N)$ time since it visits $O(\log N)$ nodes performing $O(1)$ operations in each node. In concurrent setting, up to $\vert P \vert$ other processes can be inserting their keys to the tree concurrently with \texttt{Op}, thus, at the moment of \texttt{Op.Timestamp} the size of the tree will not exceed $N + \vert P \vert$, therefore the amortized number of nodes \texttt{Op} will traverse is $O(\log N + \vert P \vert)$ (since the tree is balanced).

    In each node \texttt{v} no more than $\vert P \vert$ descriptors will be located closer to the head of \texttt{v.Queue} than the descriptor of our operation \texttt{Op}. Each operation takes $O(1)$ amortized time to execute (the rebuilding takes $O(1)$ amortized time as stated e.g., in~\cite{mehlhorn1993dynamic}), thus, \texttt{Op} takes $O(\vert P \vert)$ amortized time to finish its execution in each node.

    Therefore, amortized \texttt{Op} execution cost is $O((\log N + \vert P \vert) \cdot \vert P \vert)$.
\end{proof}

\begin{theorem}
    When the workload is uniform (i.e., each data item is equally likely to be queried)  \texttt{insert} and \texttt{remove} take $O(\log N + \vert P \vert)$ amortized time on our concurrent binary search tree with rebuilding.
\end{theorem}

\begin{proof}
    Consider the size of the root operation queue. Since there exist up to $\vert P \vert$ processes executing operations concurrently, the size of root operation queue is $O(\vert P \vert)$.

    Let us see, in which nodes these operations will continue their execution. Since each data item is equally likely to be queried, approximately half descriptors continues their execution in \texttt{root.Left} node, and the other half continues their execution in \texttt{root.Right} node. Therefore, the expected size of operation queue in each node of the second tree level is $O(\frac{\vert P \vert}{2})$.

    Following the same reasoning, the expected size of operation queue in each node of the third tree level is $O \left(\frac{\vert P \vert}{2^2}\right) = O \left(\frac{\vert P \vert}{4}\right)$ and the expected size of operation queue in each node of the $k$-th level of the tree is $O\left(\frac{\vert P \vert}{2^{k-1}}\right)$.

    Since the tree is balanced, the operation traverses $O(\log N + \vert P \vert)$ nodes. The expected amortized number of operations performed in $i$-th node is $O\left(\max\left( \frac{\vert P \vert}{2^{k-1}}, 1\right) \right)$ since the amortized cost of executing a single operation in a node is $O(1)$ (of course, in each node we perform at least $O(1)$ operations).

    Therefore, the total expected amortized cost of performing an operation is $O \left(\sum\limits_{k=1}^{\log N + \vert P \vert} \max \left( \frac{\vert P \vert}{2^{k-1}}, 1 \right) \right) = O(\log N + \vert P \vert)$.
\end{proof}

\section{Experiments}
According to the framework described in Section~\ref{algorithm-general-chapter}, we implemented a concurrent balanced binary search tree that supports \texttt{insert}, \texttt{remove}, \texttt{contains}, and \texttt{count} queries.
The code is written in Kotlin.
%and is available here~\href{}. \ik{TODO: insert ref.} \va{we cannot put a reference on github}

We decided to test our data structure only against the concurrent persistent tree presented in~\cite{aksenov2023unexpected}, since it is the only available data structure that supports asymptotically efficient range queries (e.g., can execute \texttt{count} queries in logarithmic time). %We understand that this simple persistent tree performs read operations much faster than our heavy-synchronization tree. Because of that we test these two data structures only on three simple workloads.

We test the implementations on the following workloads: 1)~a read-heavy workload that runs contains operations; 2)~an insert-delete workload with half insertions and half deletions on a random keys drawn from a range so that each operation is successful with a probability of approximately $0.5$; 3)~a successful-insert workload where we insert a random key from a very wide range (from $-2^{63}$ to $2^{63} - 1$) so that all insertions are successful with the very high probability. We consider these experiments more as preliminary rather than the full-detailed ones.

All our experiments are performed on Intel Gold 6240R with 24 cores. We decide to run on one socket due to the heavy load on the memory by our search tree. The plots show the throughput of the data structures, i.e., the number of operations in $10$ seconds. Each point on the plots is obtained as an average of $5$ separate runs. The blue lines are for our data structure, and the orange lines are for the persistent tree.

\begin{figure}[H]
    \centering
    \includegraphics[width=\linewidth]{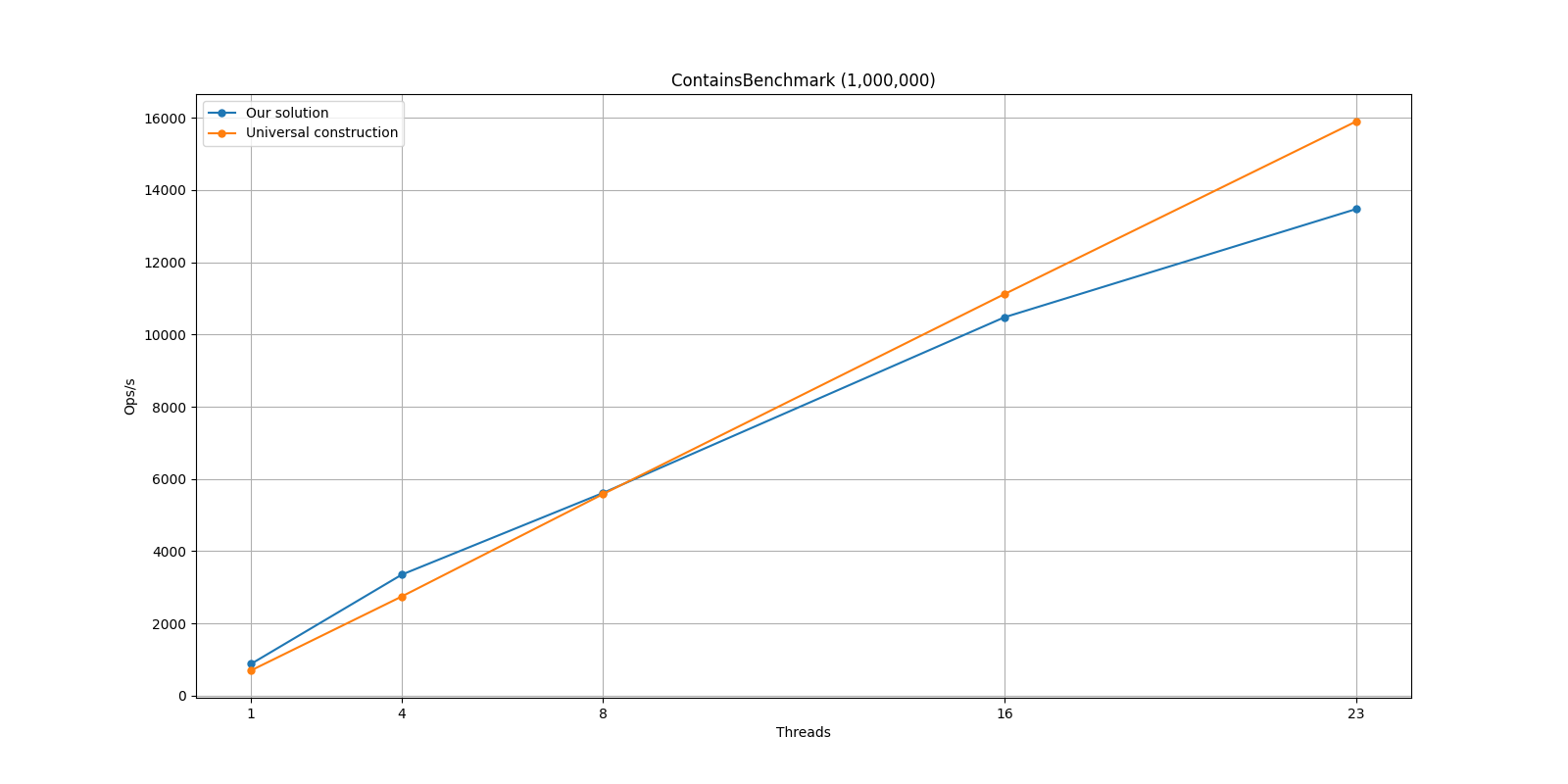}
    \caption{Contains Benchmark.}
    \label{fig:contains}
\end{figure}

\textbf{Contains Benchmark.}
We fix the key range as $[1, 2 \cdot 10^6]$. At first, we initialize a data structure~--- each element from the range is inserted with the probability $1/2$. Then, we start $T$ threads. Each thread for $10$ seconds searches for a key taken uniformly at random from the range. As shown in Figure~\ref{fig:contains}, our data structure does not have a large overhead for contains operations.

\begin{figure}
%\begin{minipage}[c]{0.46\linewidth}
    \centering
    \includegraphics[width=\linewidth]{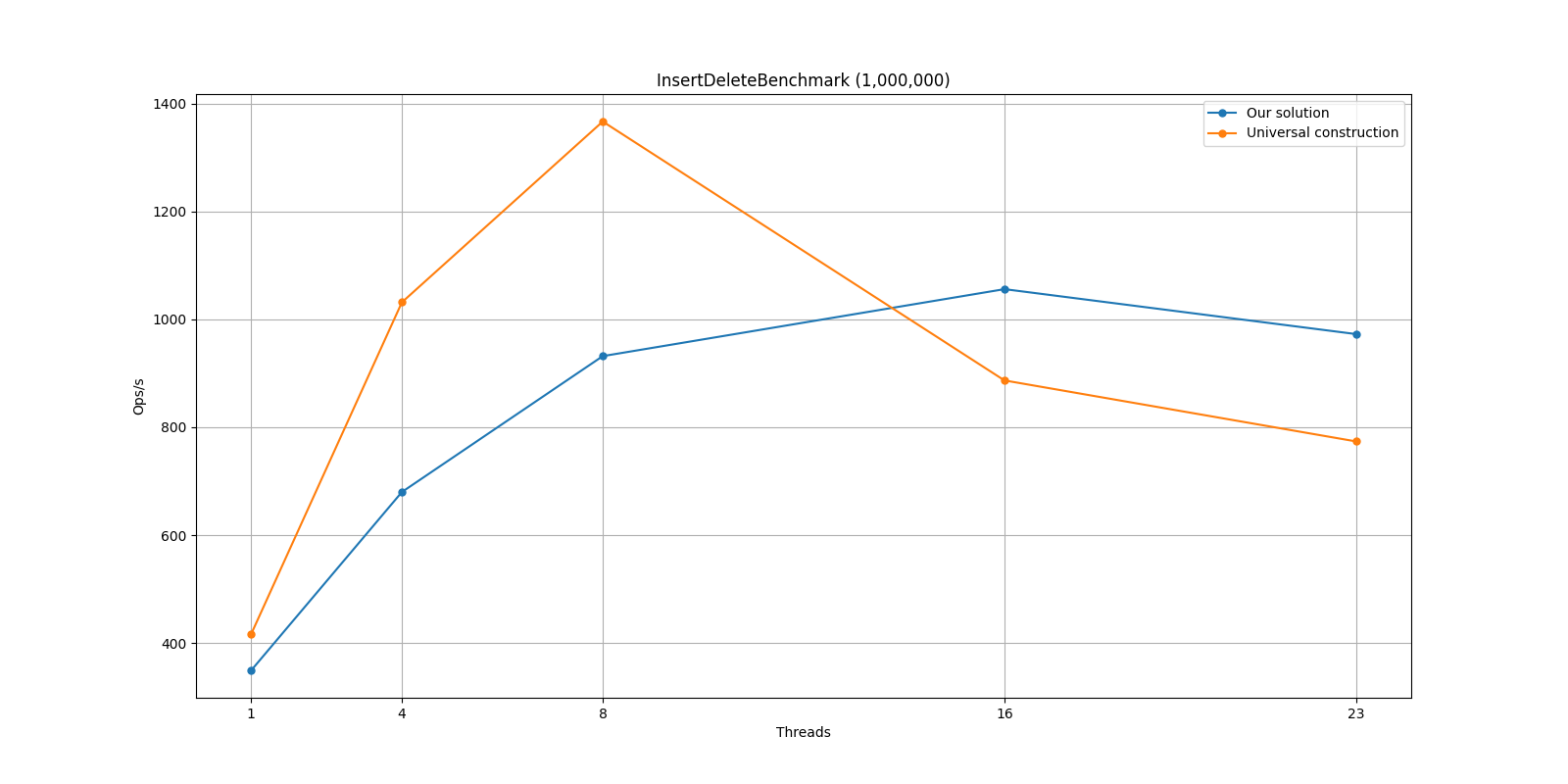}
    \caption{Insert-Delete Benchmark.}
    \label{fig:insert-delete}
\end{figure}

\textbf{Insert-Delete Benchmark.} 
We fix the key range as $[1, 2 \cdot 10^6]$. At first, we initialize a data structure~--- each element from the range is inserted with the probability $1/2$. Then, we start $T$ threads. Each thread for $10$ seconds chooses the operation (insert/delete) uniformly at random and the argument uniformly at random from the range. As shown in Figure~\ref{fig:insert-delete}, our data structure starts worse due to the larger overhead, but it works under contention better than the persistent tree.

\begin{figure}
    \centering
    \includegraphics[width=\linewidth]{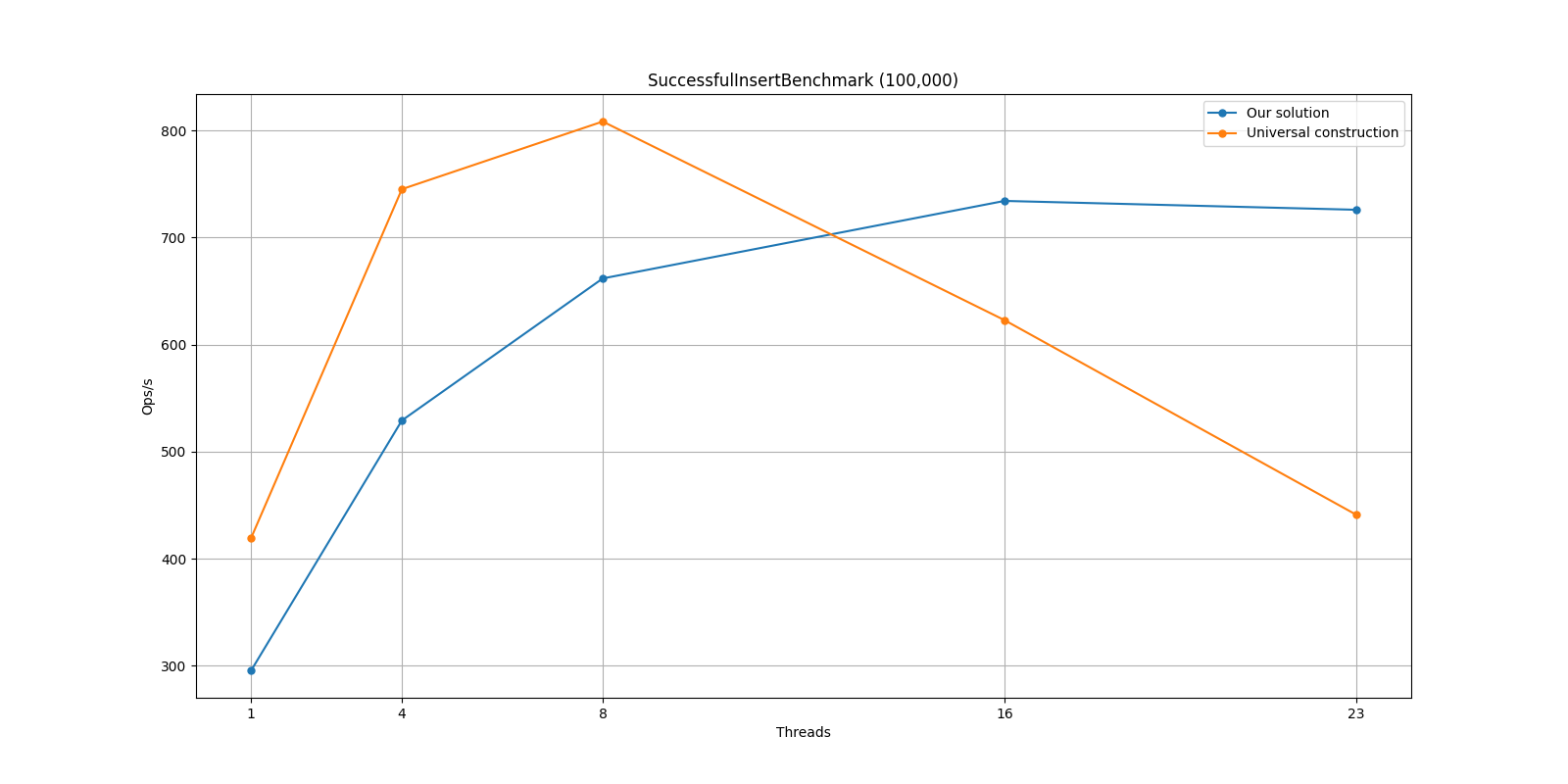}
    \caption{Successful-Insert Benchmark.}
    \label{fig:successful-insert}
%\end{minipage}
\end{figure}

\textbf{Successful-Insert Benchmark.}
We initialize a data structure with $10^6$ random integer elements. Then, we start $T$ threads. Each thread for $10$ seconds inserts random integers. With the very high probability each insertion is successful which affects the persistent tree very much. As shown in Figure~\ref{fig:successful-insert}, our data structure starts worse due to the larger overhead, but it works under contention better than persistent tree.

\textbf{Outcome.} Our experiments show that our data structure works better than the only existing solution with aggregate range queries on update-heavy workloads and has a small overhead on contains operations while supporting efficient aggregate range queries.

%\end{minipage}
%\hspace{0.05\linewidth}
%\begin{minipage}[c]{0.46\linewidth}

\section{Conclusion}
We present an approach to obtain concurrent trees with efficient aggregate range queries in a wait-free manner. Our practical results validate our performance and scalability claims. 
We propose a number of avenues for future work. First, we can make the rebuilding collaborative~\cite{brown2020non}, i.e., make different processes work together to rebuild a single subtree.
Then, in order to achieve pure $O(\log n)$ complexity, instead of the amortized one, we can use another rebuilding strategy~--- the top-down rebuilding from the chromatic tree~\cite{brown2014general}.
Another interesting question is how to decrease the number of allocations~--- now we use too many memory.
Finally, it would be good to implement other tree data structures, e.g., quad-trees or tries.

\bibliographystyle{plainurl}% the mandatory bibstyle
\bibliography{references.bib}

\appendix

\section{Efficient sequential algorithm for range queries}
\label{sequential-algorithm}

Many range queries, especially the aggregating ones, can be executed in sub-linear (e.g. logarithmic) time. Consider, an example of such range query: \newline \texttt{count(Set, min, max) = $\mid \{$ x $\in$ Set : min $\leq$ x $\leq$ max $\} \mid$}~--- the number of keys, located in the range \texttt{[min; max]}. It can be calculated in $O(\log N)$ time on binary search trees (where $N$ is the number of keys in the set), using the following algorithm.

\subsection{Tree structure}

Let us begin with a couple of definitions:

\begin{definition}
A node is a \emph{leaf} if it has no children.
\end{definition}

\begin{definition}
A node is an \emph{internal} node if it is not a leaf.
\end{definition}

\begin{figure}[H]
     \centering
     \begin{subfigure}[b]{0.55\linewidth}
          \centering
          \includegraphics[width=\linewidth]{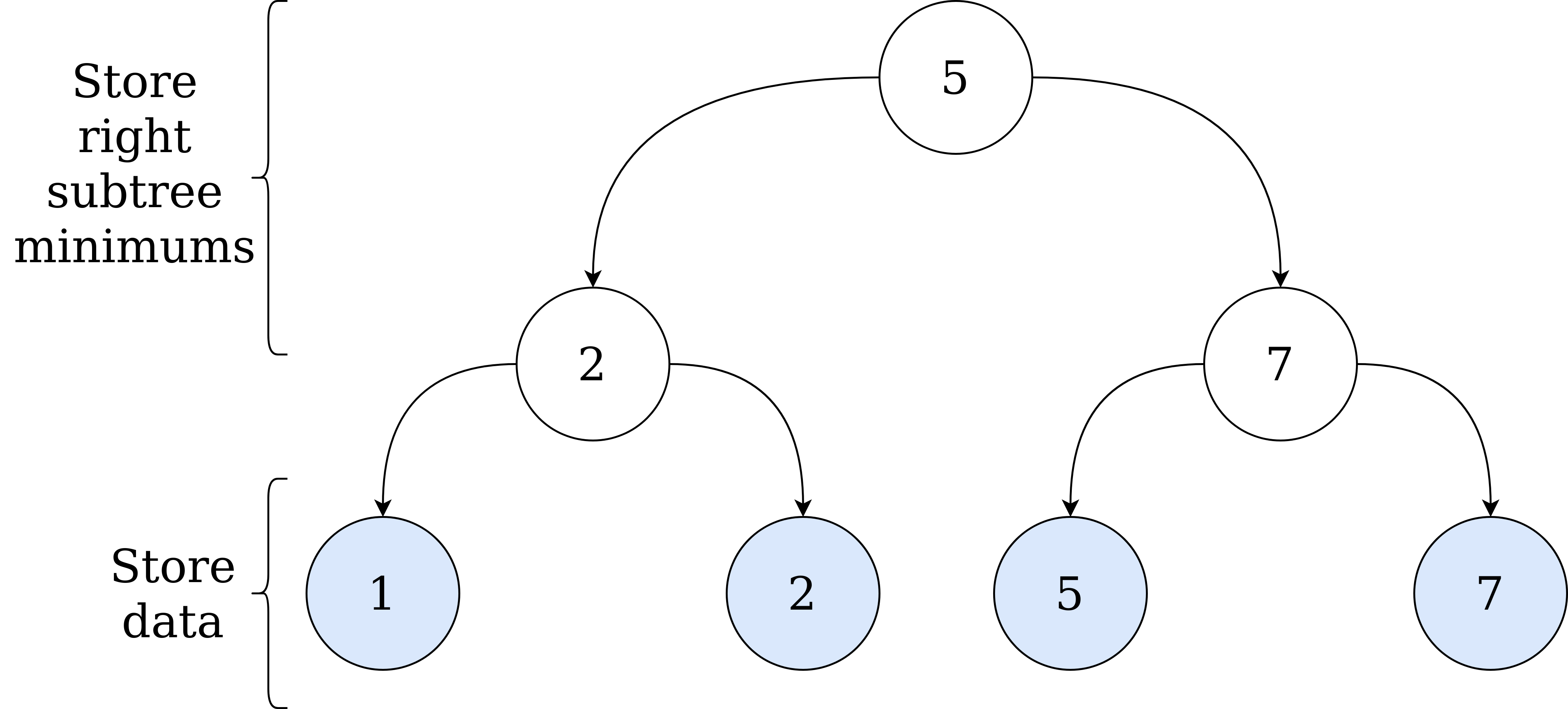}
          \caption{External binary search tree}
          \label{external-tree-pic}
     \end{subfigure}
     \hfill
     \begin{subfigure}[b]{0.35\linewidth}
          \centering
          \includegraphics[width=\linewidth]{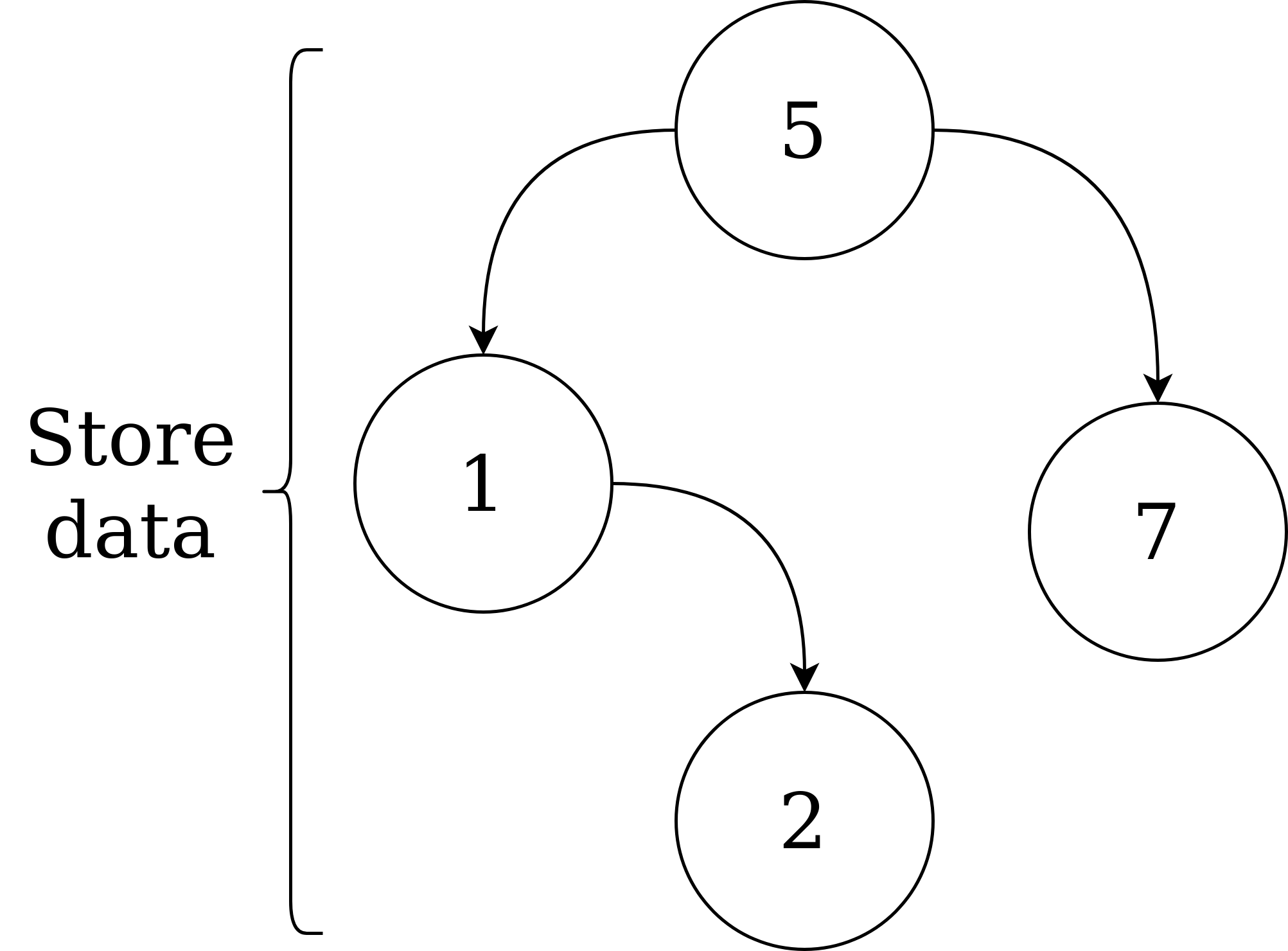}
          \caption{Internal binary search tree}
          \label{internal-tree-pic}
     \end{subfigure}
    \caption{Different types of search trees}
    \label{internal-external-trees-pic}
\end{figure}

\begin{definition}
\emph{External} binary search tree (Fig.~\ref{external-tree-pic}) is a binary search tree, in which keys are stored only in leaf nodes. In contrast, internal nodes store only auxiliary information, used for query routing (e.g., the minimal key, that might be located in the right subtree).
\end{definition}

\begin{definition}
\emph{Internal} binary search tree (Fig.~\ref{internal-tree-pic}) is a binary search tree, in which keys are stored in both leaf nodes and in internal nodes.
\end{definition}

\begin{figure}[H]
  \centering
  \caption{Using \texttt{Right\_Subtree\_Min} for query routing}
  \label{right-subtree-min-pic}
  \includegraphics[width=0.5\linewidth]{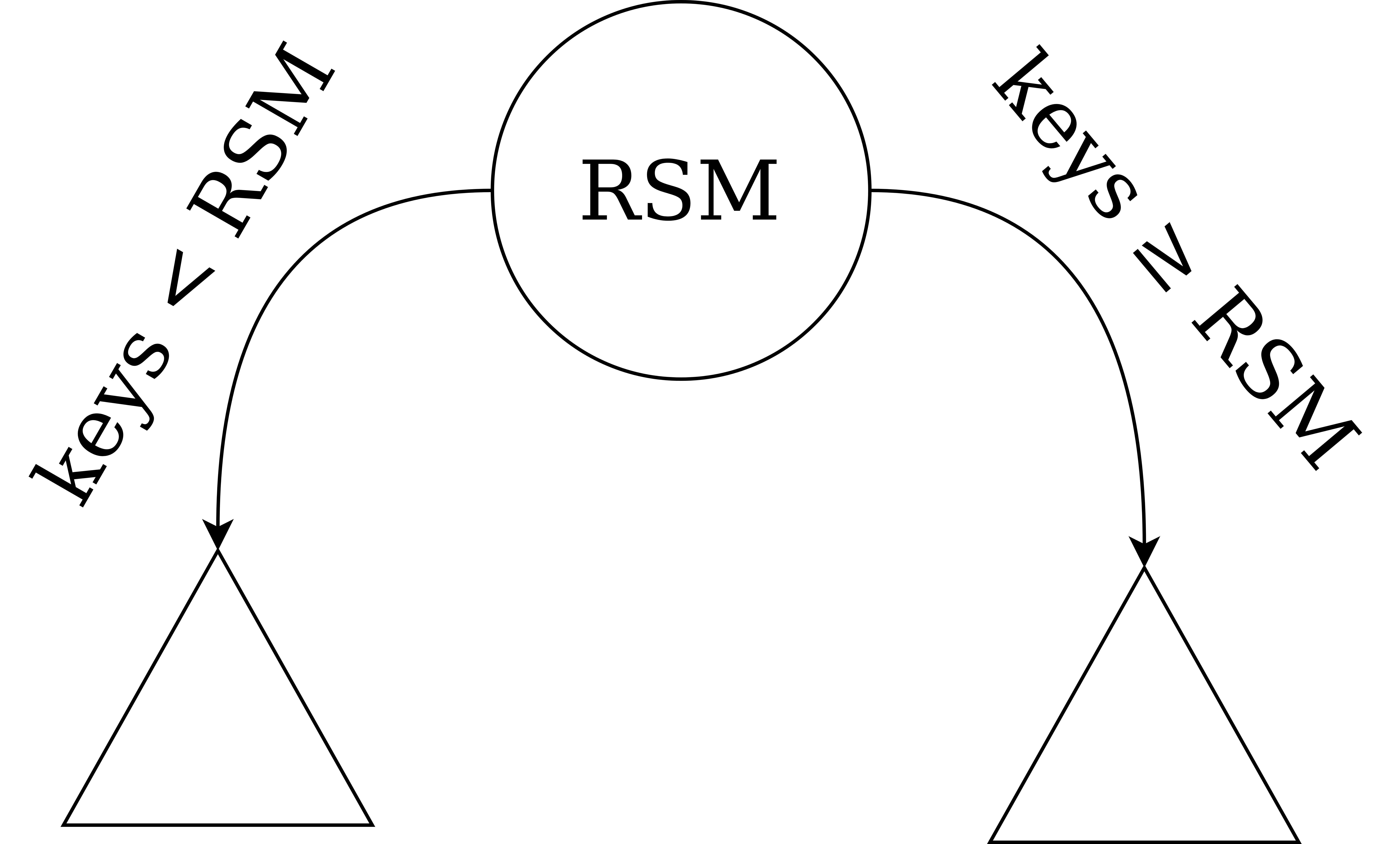}
\end{figure}

To explain how to implement the \texttt{count} query, we consider external binary search trees. Each internal node will store \texttt{Right\_Subtree\_Min}~--- the minimal key, that might be located in the right subtree. All keys less than \texttt{Right\_Subtree\_Min} should be stored in the left subtree, and, thus, all scalar queries (\texttt{insert}, \texttt{remove} and \texttt{contains}) on such keys are redirected to the left subtree. Similarly, all keys greater than or equal to \texttt{Right\_Subtree\_Min} should be stored in the right subtree, and, thus, all scalar queries on such keys are redirected to the right subtree (Fig.~\ref{right-subtree-min-pic}).

\begin{figure}[H]
  \centering
  \caption{Maintaining subtree sizes on node insertion}
  \label{insert-sizes-pic}
  \includegraphics[width=\linewidth]{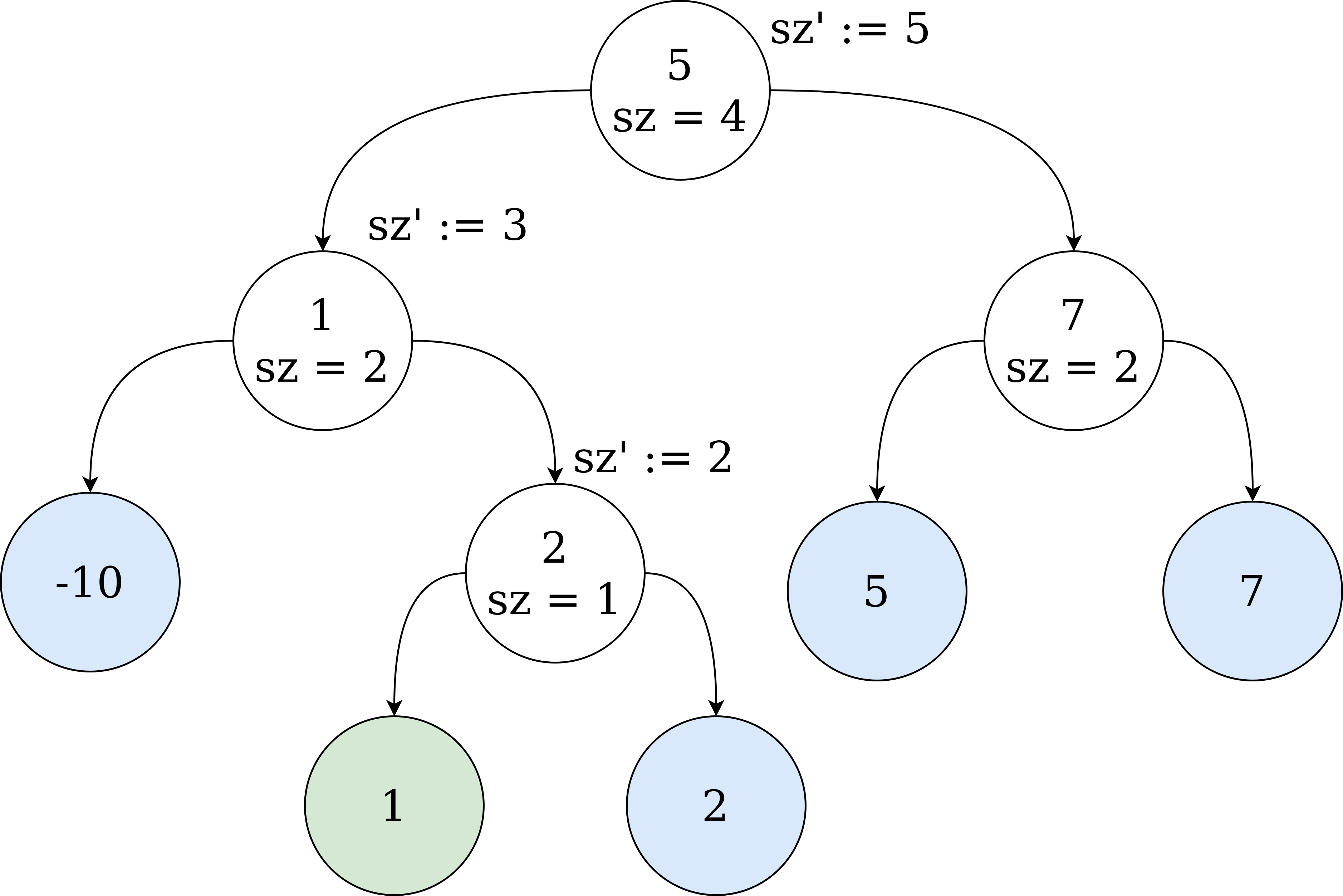}
\end{figure}

\begin{figure}[H]
  \centering
  \caption{Maintaining subtree sizes on node removal}
  \label{remove-sizes-pic}
  \includegraphics[width=\linewidth]{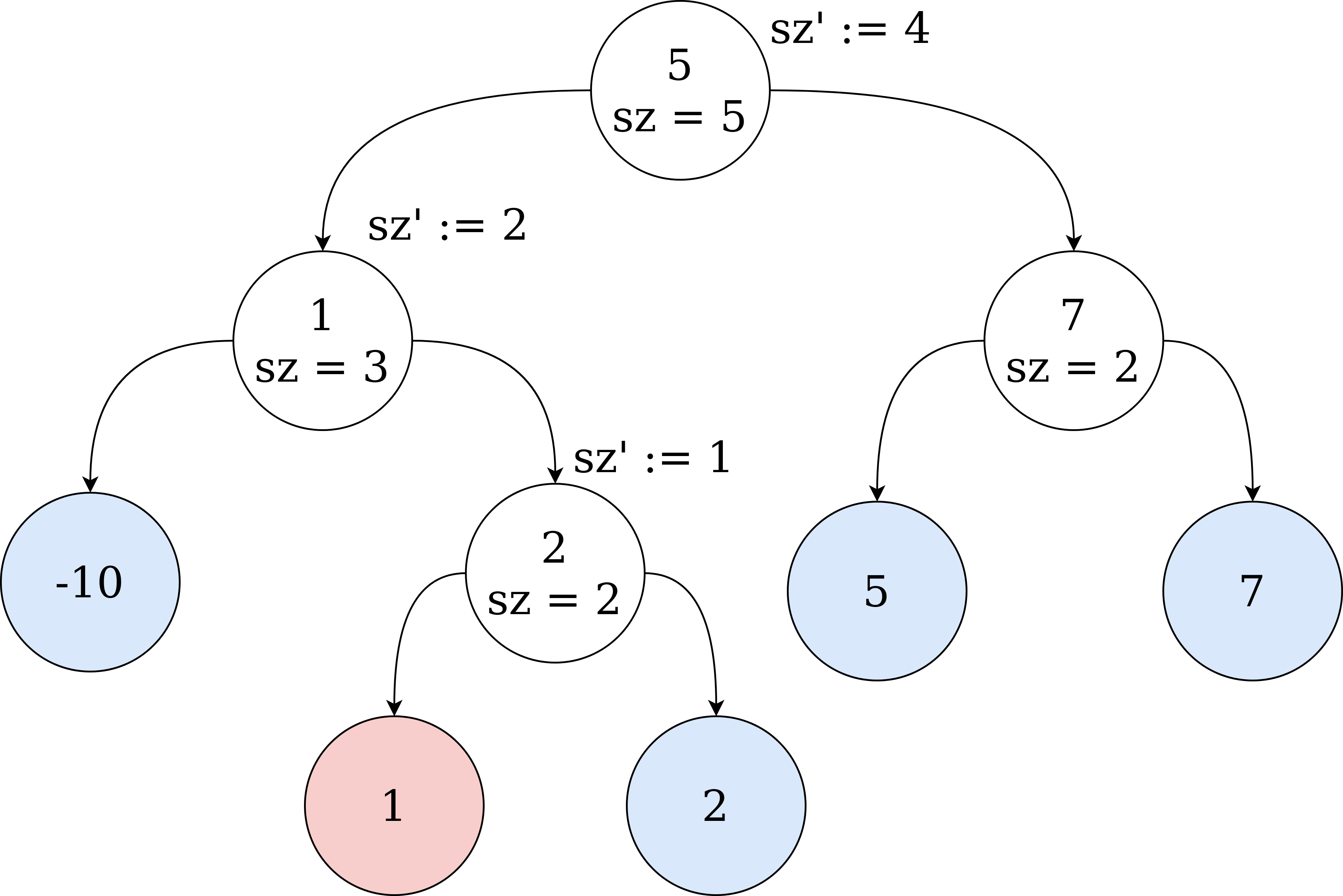}
\end{figure}

Moreover, each internal node will store the size of that node's subtree~--- i.e., the number of keys in that node's subtree. Of course, that information should be properly maintained: 
\begin{itemize}
    \item When inserting new key \texttt{k} to the tree, increase by one subtree sizes of each node on the path from the root to the leaf, storing key \texttt{k} (Fig.~\ref{insert-sizes-pic}).
    
    \item When removing key \texttt{k} from the tree, decrease by one subtree sizes of each node on the path from the root to the leaf, storing key \texttt{k} (Fig.~\ref{remove-sizes-pic}).
\end{itemize}

\begin{definition}
We call additional information, stored in tree nodes and required for fast range queries execution, \emph{augmentation values}.

For example, subtree sizes are augmentation values, required for asymptotically optimal execution of the \texttt{count} range query.
\end{definition}

Note, that different range queries may require different augmentations in order to be executed asymptotically optimal. In Chapter~\ref{applicability-chapter} we shall describe augmentations, required for fast execution of different range queries.

\subsection{Executing the \texttt{count} query asymptotically optimal}

To implement the \texttt{count} query in an asymptotically optimal way, we present the following three functions:

\begin{itemize}
    \item \texttt{count\_both\_borders(node, min, max)}~--- returns the number of keys in \texttt{node} subtree, that are located in the range \texttt{[min; max]}
    
    \item \texttt{count\_left\_border(node, min)}~--- returns the number of keys in \texttt{node} subtree, that are greater than or equal to \texttt{min}
    
    \item \texttt{count\_right\_border(node, max)}~--- returns the number of keys in \texttt{node} subtree, that are less than or equal to \texttt{max}
\end{itemize}

Trivially, \texttt{count(Set, min, max) = count\_both\_borders(Set.Root, min, max)}.

Let us begin with defining \texttt{count\_both\_borders(node, min, max)} procedure recursively:

\begin{itemize}
    \item If \texttt{node} is a leaf, we check whether \texttt{min $\leq$ node.Key $\leq$ max} holds. If so, we return \texttt{1}, otherwise, we return \texttt{0}.

    \item If \texttt{min $\geq$ node.Right\_Subtree\_Min}, then all keys from the left subtree are less than \texttt{min} (since for all such keys \texttt{Key < node.Right\_Subtree\_Min} holds, as guaranteed by the tree structure). Thus, all the required keys are located in the right subtree. Therefore, we return \texttt{count\_both\_borders(node.Right, min, max)}.
    
    \item If \texttt{max $<$ node.Right\_Subtree\_Min}, then all keys from the right subtree are greater than \texttt{max}. Thus, all the required keys are located in the left subtree. Therefore, we return \texttt{count\_both\_borders(node.Left, min, max)}.
    
    \item Otherwise, \texttt{min $<$ node.Right\_Subtree\_Min $\leq$ max}. In that case, some satisfying keys may be located in the left subtree, and some of them may be located in the right subtree. Thus, we return \texttt{count\_both\_borders(node.Left, min, node.Right\_Subtree\_Min) + count\_both\_borders(node.Right, node.Right\_Subtree\_Min, max)}. In that case, we call \texttt{node} with such a condition a \emph{fork node}.
    
    Note, that the tree structure guarantees, that all keys in the left subtree are already less than \texttt{node.Right\_Subtree\_Min} and all keys in the right subtree are already greater than or equal to \texttt{node.Right\_Subtree\_Min}. Thus, we do not need to check, that keys in the left subtree are $\leq$ \texttt{node.Right\_Subtree\_Min} and that keys in the right subtree are $\geq$ \texttt{node.Right\_Subtree\_Min}~--- these inequations are guaranteed to be true by the tree structure itself. Thus, we return \texttt{count\_left\_border(node.Left, min) + count\_right\_borders(node.Right, max)}.
\end{itemize}

Now, we shall define \texttt{count\_left\_border(node, min)}:

\begin{itemize}
    \item If \texttt{node} is a leaf, we check whether \texttt{node.Key $\geq$ min} holds. If so, we return \texttt{1}, otherwise, we return \texttt{0}.
    
    \item If \texttt{min $\geq$ node.Right\_Subtree\_Min}, then all keys from the left subtree are less than \texttt{min}. Thus, all the required keys are located in the right subtree. Therefore, we return \texttt{count\_left\_border(node.Right, min)}.
    
    \item Otherwise, \texttt{min $<$ node.Right\_Subtree\_Min}. In that case, all the keys from the right subtree are greater than or equal to \texttt{min}. Thus, we should count all keys from the right subtree plus some keys from the left subtree. Therefore, the answer is \texttt{get\_size(node.Right) + count\_left\_border(node.Left, min)}.
    
    Size of the right subtree can be calculated easily:
    
    \begin{itemize}
        \item If \texttt{node.Right} is a leaf, the size of the right subtree is \texttt{1};
        
        \item Otherwise, \texttt{node.Right} is an internal node~--- in that case the size of the right subtree is \texttt{node.Right.Size};
    \end{itemize}
\end{itemize}

We can define \texttt{count\_right\_border(node, max)} in the same manner:

\begin{itemize}
    \item If \texttt{node} is a leaf, we check whether \texttt{node.Key $\leq$ max} holds. If so, we return \texttt{1}, otherwise, we return \texttt{0}.
    
    \item If \texttt{max $<$ node.Right\_Subtree\_Min}, then all keys from the right subtree are greater than \texttt{max}. Thus, all the required keys are located in the left subtree. Therefore, we return \texttt{count\_right\_border(node.Left, max)}.
    
    \item Otherwise, \texttt{max $\geq$ node.Right\_Subtree\_Min}. In that case, all keys from the left subtree are less than \texttt{max}. Thus, we should count all keys from the left subtree plus some keys from the right subtree. Therefore, the answer is \texttt{get\_size(node.Left) + count\_right\_border(node.Right, max)}. The size of the left subtree can be calculated similarly to the previous case.
\end{itemize}

We show how to implement the algorithm in Listing~\ref{count-sequential-listing} \footnote{In all subsequent pseudocode listings we denote shared objects (including names of fields, that may be accessed by multiple processes) in \texttt{Upper\_Snake\_Case}; class names in \texttt{CamelCase}; local variables in \texttt{lower\_snake\_case}; functions in \texttt{lower\_snake\_case}; Creation of a new variable is denoted by \texttt{variable\_name := initial\_value} syntax; Assigning a new value to the existing variable is denoted by \texttt{variable\_name $\leftarrow$ new\_value} syntax;}.

\renewcommand{\lstlistingname}{Listing}
\begin{lstlisting}[caption={Implementation of the \texttt{count} range query},label={count-sequential-listing},escapeinside={(*}{*)}, captionpos=b]
fun count_both_borders(node, min, max):
    case node of
    | EmptyNode (*$\rightarrow$*)
        /* 
        EmptyNode is a dummy node that contains
        neither key nor children.
        We can use it to represent an empty set,
        for example
        */
        return 0
    | LeafNode (*$\rightarrow$*)
        if min (*$\leq$*) node.Key (*$\leq$*) max:
            return 1
        else:
            return 0
    | InnerNode (*$\rightarrow$*)
        if min (*$\geq$*) node.Right_Subtree_Min:
            return count_both_borders(
                     node.Right, min, max)
        elif max < node.Right_Subtree_Min:
            return count_both_borders(
                     node.Left, min, max)
        else:
            return
                count_left_border(node.Left, min) +
                count_right_border(node.Right, max)
    
fun get_size(node):
    case node of
    | EmptyNode (*$\rightarrow$*)
        return 0    
    | LeafNode (*$\rightarrow$*)
        return 1
    | InnerNode (*$\rightarrow$*)
        return node.Size   
                   
fun count_left_border(node, min):
    case node of
    | EmptyNode (*$\rightarrow$*)
        return 0
    | LeafNode (*$\rightarrow$*)
        if node.Key (*$\geq$*) min:
            return 1
        else:
            return 0
    | InnerNode (*$\rightarrow$*)
        if min (*$\geq$*) node.Right_Subtree_Min:
            return count_left_border(node.Right, min)
        else:
            return get_size(node.Right) + 
                   count_left_border(node.Left, min)
            
fun count_right_border(node, max):
    case node of
    | EmptyNode (*$\rightarrow$*)
        return 0
    | LeafNode (*$\rightarrow$*)
        if node.Key (*$\leq$*) max:
            return 1
        else:
            return 0
    | InnerNode (*$\rightarrow$*)
        if max < node.Right_Subtree_Min:
            return count_right_border(node.Left, max)
        else:
            return get_size(node.Left) + 
                   count_right_border(node.Right, max)
\end{lstlisting}

\subsection{\texttt{count} query time complexity}

\begin{theorem}
The time complexity of the \texttt{count} query is $O(height)$.
\end{theorem}
\begin{proof}
We state that both \texttt{count\_left\_border} and \texttt{count\_right\_border} work in $O(height)$ time. Indeed, on each tree level both these procedures visit only one node per level, performing $O(1)$ operations in each visited node.

Let us now switch to proving the time complexity of \texttt{count\_both\_borders}. At upper tree levels (higher than the \emph{fork node}) it visits one node per level performing $O(1)$ operations in each visited node, giving $O(height)$ time at upper levels. 

At one of the nodes (the \emph{fork node}) the execution may fork: we shall call \texttt{count\_left\_border} on the left subtree and \texttt{count\_right\_border} on the right subtree. Note, that the execution can fork at most once and both called procedures have $O(height)$ time complexity. Thus, at lower tree levels the procedure also has $O(height) + O(height) = O(height)$ time complexity. Therefore, the total time complexity of the procedure is $O(height)$ (Fig~\ref{sequential-time-png}).

\begin{figure}[H]
  \centering
  \caption{Time complexity of the \texttt{count\_both\_borders} procedure}
  \label{sequential-time-png}
  \includegraphics[width=\linewidth]{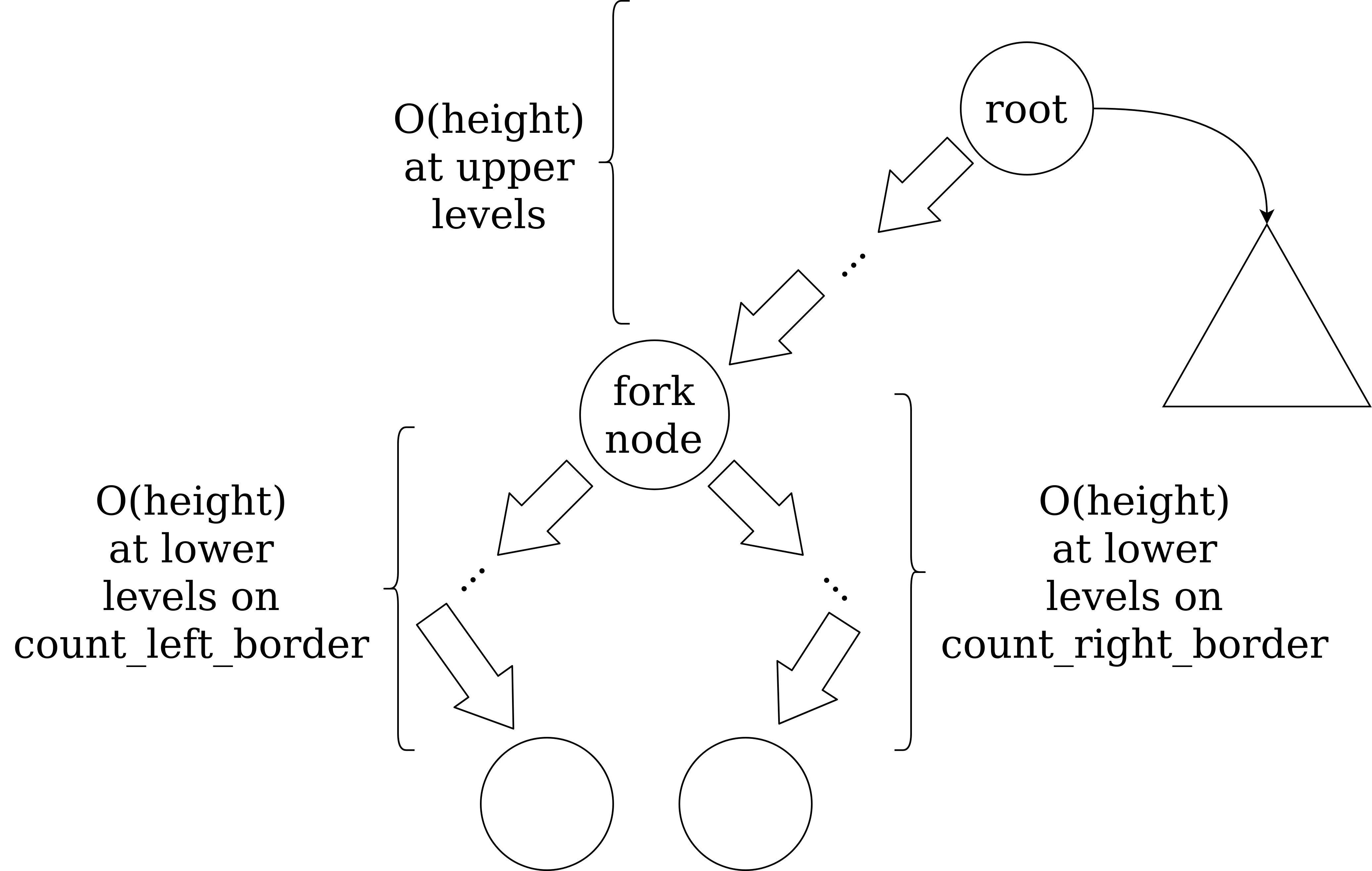}
\end{figure}
\end{proof}

Suppose we use balanced binary search trees with $height \in O(\log N)$ where $N$ is the size of the tree. Thus, the \texttt{count} query is executed in $O(\log N)$ time.

\subsection{Proof of the descriptor timestamp monotony}
\label{timestamp-monotony-proof-section}

\begin{theorem}
In each queue, operation timestamps form a monotonically increasing sequence. More formally, if at any moment we traverse any queue \texttt{Q} from the head to the tail and obtain $\texttt{t}_{\texttt{1}}, \texttt{t}_{\texttt{2}}, \ldots \texttt{t}_{\texttt{n}}$~--- a sequence of timestamps of descriptors, located in \texttt{Q}, then $\texttt{t}_{\texttt{1}} < \texttt{t}_{\texttt{2}} < \ldots < \texttt{t}_{\texttt{n}}$ will hold.
\end{theorem}

\begin{proof}
We prove the theorem by the induction on the tree structure. As the induction basis, we will show that the statement holds for the tree root. As the induction step, we will prove that, given that the statement holds for some node \texttt{pv}, the statement holds for \texttt{v}~--- an arbitrary child of \texttt{pv}. Thus, the statement is guaranteed to hold for each tree node.

\begin{itemize}
    \item As requested in Section~\ref{main-invariant-chapter} and as explained in Section~\ref{queue-section}, the root queue provides timestamp allocation mechanism with the following guarantees: if descriptor of operation \texttt{A} is inserted to the root queue before descriptor of operation \texttt{B}, then \texttt{timestamp(A) < timestamp(B)} holds. Thus, the induction base is proven.
    
    \item Consider non-root node \texttt{v} and its parent \texttt{pv}. According to the induction assumption, the statement holds for \texttt{pv}. Thus, at \texttt{pv} queue descriptor timestamps form a monotonically increasing sequence: $\texttt{t}_{\texttt{1}} < \texttt{t}_{\texttt{2}} < \ldots < \texttt{t}_{\texttt{n}}$. Consider descriptors $\texttt{D}_{\texttt{i}}$ and $\texttt{D}_{\texttt{j}}$ (Fig.~\ref{timstamps-increasing-proof-pic}), such that:
    
    \begin{itemize}
        \item Both $\texttt{D}_{\texttt{i}}$ and $\texttt{D}_{\texttt{j}}$ should continue their execution at \texttt{v};
        \item \texttt{timestamp($\texttt{D}_{\texttt{i}}$) = $\texttt{t}_{\texttt{i}}$};
        \item \texttt{timestamp($\texttt{D}_{\texttt{j}}$) = $\texttt{t}_{\texttt{j}}$};
        \item $\texttt{D}_{\texttt{i}}$ is located closer to the head of \texttt{pv} queue than $ \texttt{D}_{ \texttt{j}}$ (therefore, $\texttt{D}_{\texttt{i}}$ was inserted to \texttt{pv} queue prior to $\texttt{D}_{\texttt{j}}$)~--- thus according to the induction assumption $\texttt{t}_{\texttt{i}} < \texttt{t}_{\texttt{j}}$.
    \end{itemize}
    
    \begin{figure}[H]
      \centering
      \caption{Descriptor $\texttt{D}_{\texttt{i}}$ is located closed to the head of \texttt{pv} queue than $\texttt{D}_{\texttt{j}}$, both $\texttt{D}_{\texttt{i}}$ and $\texttt{D}_{\texttt{j}}$ will continue their execution in \texttt{v} subtree}
      \label{timstamps-increasing-proof-pic}
      \includegraphics[width=\linewidth]{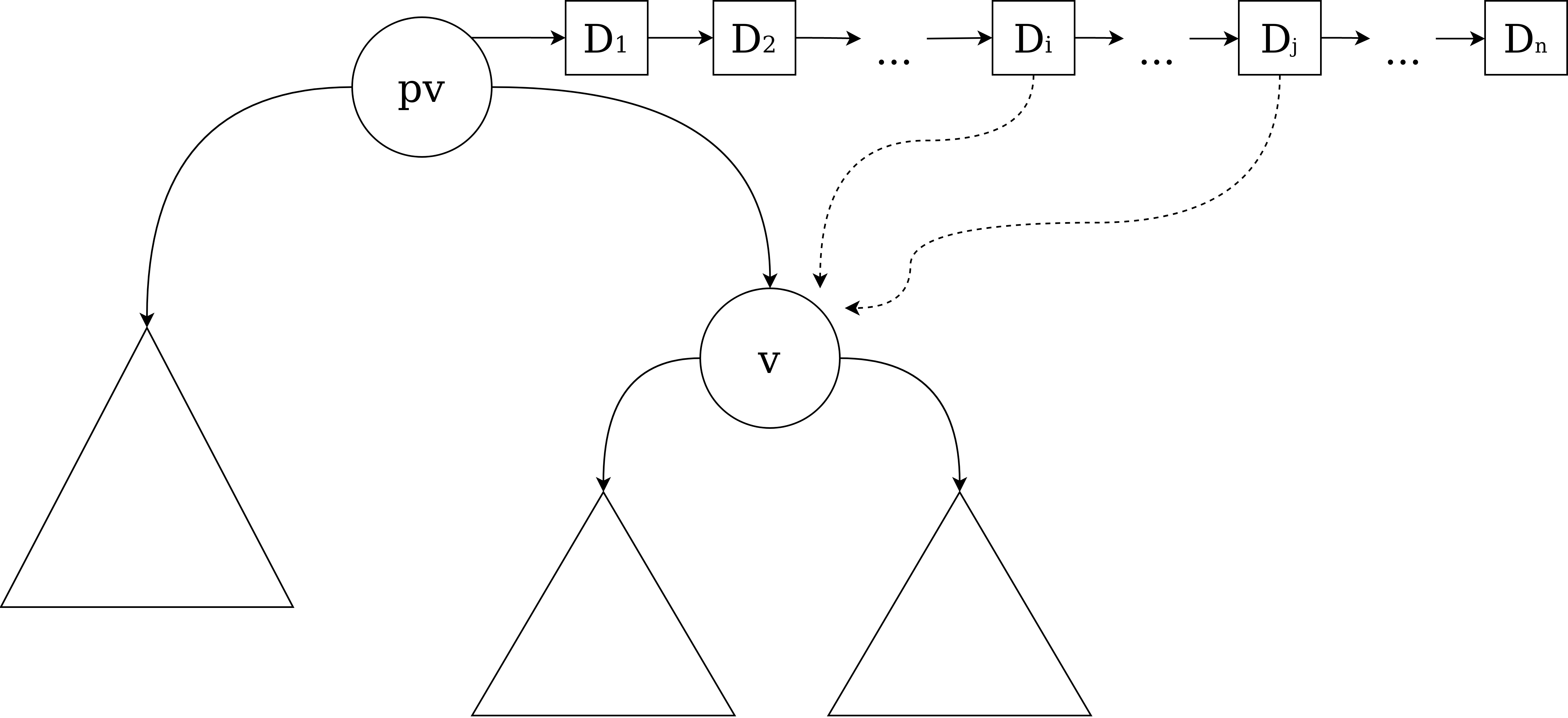}
    \end{figure}
    
    In that case, according to the algorithm, the execution of $\texttt{D}_{\texttt{j}}$ in \texttt{pv} cannot begin until the the execution of $\texttt{D}_{\texttt{i}}$ in \texttt{pv} is finished. Since the execution of $\texttt{D}_{\texttt{i}}$ in \texttt{pv} includes inserting $\texttt{D}_{\texttt{i}}$ into \texttt{v} queue, the execution of $\texttt{D}_{\texttt{j}}$ in \texttt{pv} cannot begin until $\texttt{D}_{\texttt{i}}$ is inserted into \texttt{v} queue. Thus, $\texttt{D}_{\texttt{i}}$ is inserted into \texttt{v} queue prior to $\texttt{D}_{\texttt{j}}$, thus the timestamps increasing property holds for \texttt{v}.
\end{itemize}
\end{proof}

\end{document}